\documentclass[
superscriptaddress,
preprint,
 amsmath,amssymb,
 aps,
prb,
]{revtex4-2}

\usepackage{graphicx}
\usepackage{dcolumn}
\usepackage{bm}
\usepackage{bbm}
\usepackage{hyperref}
\usepackage[mathlines]{lineno}
\usepackage{xcolor}

\begin{document}

\title{Floquet theory and computational method for the optical absorption of laser-dressed solids} 

\author{Vishal Tiwari}
\affiliation{Department of Chemistry, University of Rochester, Rochester, New York 14627, USA}

\author{Bing Gu}
\affiliation{Department of Chemistry and Department of Physics, School of Science, Westlake University, Hangzhou, Zhejiang  310030, China}

\author{Ignacio Franco}
\email{ignacio.franco@rochester.edu}
\affiliation{Department of Chemistry, University of Rochester, Rochester, New York 14627, USA}
\affiliation{Department of Physics, University of Rochester, Rochester, New York 14627, USA}

\date{\today}

\begin{abstract}
Recent advances in laser technology now enable engineering the electronic structure of matter through strong light-matter interactions.  However, the effective physico-chemical properties of these laser-dressed non-equilibrium materials are not well understood. 
Here we develop a general theory that now enables modeling and interpreting the linear optical absorption of solids that are dressed by light of arbitrary strength and photon energy. The theory applies to any crystalline solid, including dielectrics, semiconductors, semimetals and quantum materials. In the theory, the dressing of Bloch electrons by the driving laser is treated exactly using Floquet theory. The effective optical properties of this laser-dressed material are probed through a weak laser whose effects are captured to first order in perturbation theory. Remarkably, in this  non-equilibrium system the time- and space-periodic Floquet-Bloch modes play the role of the pristine eigenstates of matter as the optical absorption is seen to emerge from transitions among them. 
We implement the theoretical framework into a code (\emph{FloqticS}: Floquet optics in Solids) which is made available through \texttt{GitHub} and can be interfaced with first-principle based electronic-structure computational packages. To isolate the emergent phenomenology, we performed computations in a model solid with a cosine-shaped lattice potential driven by strong non-resonant light. The computations recover the dynamical Franz-Keldysh effect and identify novel dramatic changes in the optical absorption upon increasing the amplitude of the driving laser. The Floquet replicas open absorption sidebands separated by integer multiples of the drive photon energy. The hybridization of the Floquet-Bloch modes, create intense low-frequency absorption and stimulated emissions, and dips in the absorption spectrum. We assign these emerging effects as purely-optical tell-tale signatures of the Floquet-Bloch modes.  These advances can be used to model, control, and characterize the response properties of laser-dressed materials.
\end{abstract}

\maketitle


\section{Introduction}\label{sec1}

Modern technology now enables to strongly couple light with matter opening unprecedented opportunities to generate novel light-dressed states of matter with unique physico-chemical properties. The light dressing can be exerted through lasers \cite{Kruchinin2018, Torre2021}, by coupling to metal nanoplasmonics \cite{Dombi2020}, or by introducing matter into optical cavities \cite{Feist2017, Ribeiro2018, Schlawin2022}.

In particular, the latest advances in laser technology enable the generation and control of few-cycle lasers in the IR and UV/Vis. Using such pulses, it is now possible to apply laser fields with intensities of $\sim 10^{13}-10^{14}$ W cm$^{-2}$ (amplitudes of $\sim 1-2$ V/\AA) before the emergence of dielectric breakdown. At those intensities, the incident light can dramatically distort the electronic structure of bulk matter as the strength of the light-matter interaction becomes comparable to the strength of chemical bonds, thus opening exciting opportunities to create laser-dressed materials with structure-function relations that can be very different from those observed near thermodynamic equilibrium. 
Strong field effects that are expected at such electric field amplitudes include Zener interband tunneling of electrons \cite{Zener1934}, Bloch oscillations \cite{Bloch1929, Hartmann2004}, the appearance of localized electronic states and the associated Wannier-Stark ladder in the energy spectrum \cite{Wannier1960, Shockley1972, Carlo1994, Schmidt2018}, and the response of electronic dynamics to the subcycle structure of light \cite{Franco2007, Chen2018,  GarzonRamirez2018,  Schiffrin2012}. Recent demonstrations of emerging effects in this regime include the creation of light-induced conical intersections \cite{Natan2016, Kuebel2020},  superconductivity \cite{Mitrano2016,Budden2021}, ultra-fast switching of electronic phases \cite{Stojchevska2014}, high harmonic generation \cite{Yoshikawa2017, UzanNarovlansky2022}, the Stark control of electrons at interfaces (SCELI) \cite{Antonio2023, GarzonRamirez2021, GarzonRamirez2020, GarzonRamirez2018}, ultrafast currents \cite{ Cocker2016, Huber2016, Higuchi2017, Heide2019, Ludwig2019,  Peller2020}, and logic gates that operate at the petahertz limit \cite{Boolakee2022}. 


To enhance our ability to characterize and control matter driven far from equilibrium, it is necessary to develop useful general theories of the response properties of laser-dressed matter. 
Here, we are concerned with the optical absorption properties of laser-driven solids. We envision a physical situation, as experimentally realized in Refs. \cite{Srivastava2004, Ghimire2011, Novelli2013, Lucchini2016,  Shan2021, Volkov2023, Kobayashi2023}, in which a laser drives matter far from equilibrium, while a second perturbative laser source probes its effective ability to absorb light across the electromagnetic spectrum.

The theoretical challenge is to generalize the usual theory of  optical absorption to this nonequilibrium situation where matter is constantly driven by light. New theoretical tools are needed because, in this regime, there is no stationary reference state and, since energy is no longer a conserved quantity, the increase of energy of a system from a given reference state can no longer be used as a criterion for the optical absorption. Further, the fluctuation-dissipation theorem \cite{Kubo1963} and Green-Kubo relations \cite{Green1954, Kubo1957}, that form the basis of the usual linear response theory \cite{mukamel1995}, are no longer valid since the Hamiltonian of driven matter is not time-translational invariant. In turn, fully perturbative approaches  of the response of matter to both driving and probing pulse, while possible, cannot capture the dynamics induced by the driving pulse exactly in all regimes of the laser-matter interaction. In addition, numerical approaches \cite{Otobe2016, Du2019, TancogneDejean2020, Dong2022} in which the response is computed by directly propagating the dynamics using time-dependent density functional theory framework or Green's functions  \cite{Freericks2009, Schueler2020} can quickly become computationally challenging for realistic solids.

Here we develop the theory for the linear optical absorption of solids  dressed with light of arbitrary frequency and intensity. The theory applies to crystalline solids with any band structure and dimensionality. We further  develop the computational means to integrate such a theory with standard first-principle based electronic structure codes as needed to quantify such a response in realistic systems. We invoke Floquet theory\cite{Floquet1883, Shirley1965, Sambe1973}  to exactly capture the effects of the driving laser while avoiding the computational cost of directly propagating the dynamics. In turn, the effects of the laser that probes the effective optical properties are captured to first order in perturbation theory.

Floquet theory essentially maps the time-dependent Schr\"odinger equation for a periodically driven system into an eigenvalue problem for a Floquet Hamiltonian in an extended Hilbert space called Sambe space \cite{Sambe1973}. While, strictly speaking, Floquet theory only applies to periodically driven systems (such as that provided by continuous wave laser excitation),  recent observations have demonstrated that the Floquet picture and physics remain valid even when the driving is done with pulsed light \cite{Wang2013,Wang2017,Aeschlimann2021,Park2022,Zhou2023} and, most recently, even with few-cycle laser pulses \cite{Lucchini2022, Ito2023} making the Floquet picture of general applicability.

 Specifically, we generalize our recently proposed theory of optical absorption of laser-dressed nanomaterials \cite{Gu2018} to crystalline solids and develop a simulation strategy that now enables to study extended systems. To preserve the spatial periodicity of solids, even in presence of laser fields, we include the light-matter interaction in the velocity gauge (i.e., $\hat{\mathbf{P}} \cdot \mathbf{A}$), as opposed to the length gauge in Ref. \cite{Gu2018}, enabling us to use the Bloch theorem throughout. To quantify the optical absorption in the laser-dressed system, we capture all transitions induced by the probe laser that lead to net photon absorption or emission to first order in perturbation theory. As shown, the optical absorption of nonequilibrium system is determined by a two-time correlation function of the momentum. This contrasts to the one-time correlation function that is characteristic of near-equilibrium systems, because the drive laser breaks the time-translation symmetry of the system. By invoking Floquet theory combined with the Bloch theorem, we compute this two-time correlation function without numerically propagating the system in the presence of both fields. The final expressions obtained for net absorption in laser-dressed system are reminiscent to the near-equilibrium theory of optical absorption but with the Floquet-Bloch modes --eigenstates of the Floquet Hamiltonian for spatially periodic systems-- playing the role of system eigenstates.

 We computationally implement these equation into a general code named \emph{FloqticS} (Floquet optics in solids) that takes the electronic structure of a solid as input and outputs the absorption spectrum for a given laser drive. The input characterizing the solid can be obtained either analytically for model system or from density functional theory (DFT) computational packages. The code takes advantage of the parallel diagonalization package ELPA --Eigenvalue soLvers for Petaflop Applications \cite{Marek2014}-- to efficiently solve the computational challenge of diagonalizing the Floquet-Bloch Hamiltonian. In this way, it makes it possible to model realistic solids including reciprocal space vectors over the whole Brillouin zone. 

 To isolate the emerging phenomenology, we compute the absorption spectrum of a laser-dressed solid described by a cosine-shaped lattice potential. The laser-dressed absorption spectrum naturally recovers the blue shift of the band edge and below band-gap absorption as the drive electric field strength is increased that are characteristic of the  dynamical Franz-Keldysh effect (DFKE) \cite{Jauho1996, Johnsen1998, Chin2000, Srivastava2004, Otobe2016}. However, while the DFKE is based on parabolic band models and density of states considerations, our theory can naturally incorporate other types of band structures, capture changes of level occupations due to photoexcitation, and  variations  of the optical transition momentum matrix elements across the Brillouin zone and due to laser-driving. These effects are key to describe the properties of laser-dressed solids \cite{Hawkins2015, Yue2022} and are  not simply related to density of states considerations.

Importantly, the computations  reveal three purely optical signatures of the existence of Floquet-Bloch states. First, the spectra shows multiple optical sidebands  that are energetically separated from one another by integer multiples of the drive photon energy that reflect the replicas in the Floquet-Bloch energy eigenspectrum.  Surprisingly, these replicas are visible even for the congested electronic structure of solids. Remarkably, we also observe that the \emph{hybridization} of Floquet-Bloch modes leads to the emergence of intense absorption and stimulated emissions in the low-frequency region ($<0.6$ eV) of the absorption spectrum and dips in the absorption spectra at integer multiples of the drive photon energy. Both these effects are challenging to explain without invoking the existence of Floquet-Bloch states.

 Overall, our developments show that the Floquet-Bloch modes  provide a natural basis to understand the response properties of laser-dressed solids. They play a role that is akin to the one of the pristine energy eigenstates of the system in near-equilibrium matter. In particular, the optical absorption is seen to arise from optically induced transitions between the Floquet-Bloch modes. This observation provides theoretical insights into  the growing body of evidence that shows that the Floquet-Bloch states are useful in understanding the properties of laser-dressed solids \cite{Wang2013, Flaeschner2016, Shan2021, Earl2021, Aeschlimann2021, Park2022}.
The developed theory can also be used as a starting point for future efforts to capture higher-order terms in the optical response \cite{Shan2021},  and  additional features introduced by connecting the system to a thermal environment \cite{Engelhardt2019, Mori2023}, or by many-body electron-electron and electron-nuclear interactions that can contribute to heating and the broadening of the spectral features.

 This paper is organized as follows: In Sec. \ref{sec2} we introduce the theory of non-equilibrium optical absorption for extended systems. In Sec. \ref{sec3} we discuss the computational approach  used to implement the theory. In Sec. \ref{sec4}, we discuss computations of the optical absorption of a model solid with varying drive laser parameters and provide a useful interpretation. In Sec. \ref{sec5}, we summarize our main findings and advance a qualitative physical picture to explain our observations.

\section{Theory} \label{sec2}

\subsection{Hamiltonian}\label{sec2a}

 The Hamiltonian for a solid in the presence of a probe and drive laser field in dipole approximation is
\begin{align}
\label{parts}
\hat{H}(\hat{\bm{\mathit{r}}},t)&=\hat{H}_{\textrm{LD}}(t)+\hat{H}_{\textrm{p}}(t) ,
\end{align}
where
\begin{align}
\hat{H}_{\textrm{LD}}(t)&=\frac{\hat{\mathbf{P}}^2}{2m_{e}}+V( \hat{\mathbf{r}}_{1} , \hat{\mathbf{r}}_{2} , \hdots , \hat{\mathbf{r}}_{N} )  +\frac{e\mathbf{A}_{\textrm{d}}(t)\cdot\hat{\mathbf{P}}}{m_{e}} 
\end{align}
is the many-body Hamiltonian of the laser-dressed solid and 
\begin{align}
\label{probeham}
\hat{H}_{\textrm{p}}(t)&= \frac{e\mathbf{A}_{\textrm{p}}(t)\cdot\hat{\mathbf{P}}}{m_{e}}
\end{align}
is the interaction due to the probe laser. Here, $\{ \hat{\mathbf{r}}_{1} , \hat{\mathbf{r}}_{2} , \hdots , \hat{\mathbf{r}}_{N} \}$ and $\hat{\mathbf{P}}=\{\hat{\mathbf{p}}_{1},\hat{\mathbf{p}}_{2},\hdots,\hat{\mathbf{p}}_{N}\}$ represent the position and momentum operator for the $N$-electron system respectively, $m_{e}$ is the mass of electron and $-e$ its charge. The potential $ V( \hat{\mathbf{r}}_{1} ,  \hdots, \hat{\mathbf{r}}_{N} ) = V( \hat{\mathbf{r}}_{1} + \mathbf{R} ,  \hdots, \hat{\mathbf{r}}_{N} + \mathbf{R}) $ is spatially periodic, where $\{\mathbf{R}\}$ are the primitive lattice vectors. In turn, $\mathbf{A}_{\textrm{p}}(t)$ is the vector potential due to probe laser and $\mathbf{A}_{\textrm{d}}(t)$ due to drive. Terms proportional to  $\mathbf{A}^2(t)$ are not included as they just add an overall time-dependent shift in the zero of energy with no observable consequences and can be rigorously removed from the Hamiltonian through unitary transformations (Appendix A in Ref. \cite{Hsu2006}).  

The electric field of the drive laser can be taken to be of any general time-periodic form and polarization. For simplicity in presentation here we take the drive laser electric field as $\mathbf{E}_{\textrm{d}}(t)=-\frac{d \mathbf{A}_{\textrm{d}}(t)}{d t}=E_{\textrm{d}}\cos(\Omega t)\hat{\mathbf{e}}_{\textrm{d}}$, where  $E_{\textrm{d}}$ is its amplitude, $\hbar \Omega$ its photon energy, and $\hat{\mathbf{e}}_{\textrm{d}}$ the polarization direction.   Similarly, the  electric field due to the probe laser is $\mathbf{E}_{\textrm{p}}(t)=-\frac{d \mathbf{A}_{\textrm{p}}(t)}{d t}=E_{\textrm{p}}\cos(\omega t)\hat{\mathbf{e}}_{\textrm{p}}$, where  $E_{\textrm{p}}$ is the amplitude, $\hbar \omega$ the photon energy, and $\hat{\mathbf{e}}_{\textrm{p}}$ the probe laser polarization unit vector. Thus, $\mathbf{A}_{\textrm{d}}=-\frac{E_{\textrm{d}}}{\Omega}\sin(\Omega t)\hat{\mathbf{e}}_{\textrm{d}}$ and $\mathbf{A}_{\textrm{p}}=-\frac{E_{\textrm{p}}}{\omega}\sin(\omega t)\hat{\mathbf{e}}_{\textrm{p}}$.  Note that by treating the drive and the probe laser in velocity gauge, the total Hamiltonian Eq. \eqref{parts} maintains its periodicity in space \cite{Krieger1986,Tamaya2016}. While the total Hamiltonian is not periodic in time due to the presence of probe and drive laser, the laser-dressed Hamiltonian $\hat{H}_{\textrm{LD}}(t)$ is periodic with time period $T=\frac{2\pi}{\Omega}$.

In what follows, we adopt the following notation: $u,v,r,s$ denote the pristine band index of the solid, 
 $\alpha, \beta, \gamma, \delta$ denote  the Floquet-Bloch states, and $|\Psi_{a}\rangle, |\Psi_{b}\rangle$ denote the many-body states.

In second quantization, the Hamiltonian of the laser-dressed solid is
\begin{align}
\nonumber
\hat{H}_{\textrm{LD}}(t) & = \sum_{\mathbf{k}}\sum_{u,v} \langle \psi_{u\mathbf{k}}|\mathcal{\hat{H}}_{\textrm{LD}}(t)|\psi_{v\mathbf{k}}\rangle \hat{c}^{\dagger}_{u\mathbf{k}}\hat{c}_{v\mathbf{k}} \\
\label{laserdress}
&= \sum_{\mathbf{k}}\sum_{u,v} \left( \epsilon_{u\mathbf{k}}\hat{c}^{\dagger}_{u\mathbf{k}}\hat{c}_{u\mathbf{k}} + \frac{e\mathbf{A}_{\textrm{d}}(t)}{m_{e}}\cdot \mathbf{p}_{u\mathbf{k},v\mathbf{k}} \hat{c}^{\dagger}_{u\mathbf{k}}\hat{c}_{v\mathbf{k}} \right) ,
\end{align}
where 
\begin{align}
\mathcal{\hat{H}}_{\textrm{LD}}(t)& =\frac{\hat{\mathbf{p}}^{2}}{2m_{e}}+V_{0}(\hat{\mathbf{r}})+\frac{e}{m_{e}}\mathbf{A_{\textrm{d}}}(t)\cdot \hat{\mathbf{p}} 
\end{align}
is the effective single-particle Hamiltonian of the solid as constructed from density functional theory with effective single-particle interaction potential $V_{0}(\hat{\mathbf{r}}) = V_{0} (\hat{\mathbf{r}}+\mathbf{R})$. The operator $\hat{c}^{\dagger}_{u\mathbf{k}}$ creates a single-particle in Bloch state $| \psi_{u\mathbf{k}} \rangle = \frac{1}{\sqrt{V}} e^{i \mathbf{k} \cdot \hat{\mathbf{r}}} | u \mathbf{k} \rangle$  where $u$ labels the band and  $\mathbf{k}$ the crystal momentum with band energy $\epsilon_{u\mathbf{k}}$ and $V$ is the volume of the crystal. The Bloch function $\langle \mathbf{r} |  u \mathbf{k} \rangle = \langle \mathbf{r+R} |  u \mathbf{k} \rangle $ is a periodic function with the periodicity of the lattice. The creation and annihilation operators satisfy the usual fermionic anti-commutation relations $\{\hat{c}_{u\mathbf{k}},\hat{c}_{v\mathbf{k'}}\}=\{\hat{c}^{\dagger}_{u\mathbf{k}},\hat{c}^ {\dagger}_{v\mathbf{k'}}\}=0 $ and $ \{\hat{c}^{\dagger}_{u\mathbf{k}},\hat{c}_{v\mathbf{k'}}\}=\delta_{uv}\delta_{\mathbf{k}\mathbf{k'}}$. The second term in Eq. \eqref{laserdress}  arises due to the interaction of the drive laser with the Bloch electrons. The matrix elements of the single-particle momentum operator $\hat{\mathbf{p}}$ are
\begin{align}
\nonumber
\mathbf{p}_{u\mathbf{k},v\mathbf{k'}} = \langle \psi_{u\mathbf{k}} | \hat{\mathbf{p}} | \psi_{v\mathbf{k}'}\rangle &= \frac{1}{V} \langle u \mathbf{k} | e^{-i \mathbf{k} \cdot \hat{\mathbf{r}}} \hat{\mathbf{p}} e^{ i\mathbf{k}' \cdot \hat{\mathbf{r}}} | v \mathbf{k}'\rangle \\ 
 \nonumber
 & = \frac{1}{V} \delta_{\mathbf{k}\mathbf{k'}} M \langle u \mathbf{k} | (\hat{\mathbf{p}} + \hbar \mathbf{k'})| v \mathbf{k'} \rangle_{\textrm{UC}} \\
  \label{momentumcoup}
 & = \frac{1}{V} \delta_{\mathbf{k}\mathbf{k'}} \langle u\mathbf{k} | (\hat{\mathbf{p}} + \hbar \mathbf{k'})| v \mathbf{k'} \rangle ,
\end{align}
where $M$ is the number of unit cells in the crystal and $\langle \cdots \rangle_{\textrm{UC}}$ represents an integral over the unit cell. Thus, the laser driving can only lead to vertical transitions in reciprocal space that do not change the momentum of the charge carriers \cite{Ernotte2018}.

\subsection{Optical Response in Terms of Two-Time Momentum Correlation Function}\label{sec2b}

To quantify the absorption spectrum of the laser-dressed system prepared in a many-body state $|\Psi_{a}\rangle$ at time $t_{0}$, we compute the rate of transitions induced by the probe laser
\begin{equation} 
I(\omega)=\lim_{t\rightarrow\infty}\frac{W(t,\omega)}{t-t_{0}} ,
\end{equation}
where $W(t,\omega)$ is the probability of a probe photon of frequency $\omega$ being absorbed or emitted in the laser-driven material after an interaction time interval $t-t_{0}$. Such a quantity leads to Fermi golden rule in linear response theory and has also been used to compute the absorption properties of laser-driven matter \cite{Gu2018,Mizumoto2005, Mizumoto2006}.

In what follows, it is useful to exactly decompose the total evolution operator as $\hat{U}
(t,t_{0})=\hat{U}_{\textrm{d}}(t,t_0)\hat{U}_{\textrm{p,I}}(t,t_{0})$, where 
$\hat{U}(t, t_0)$ satisfies the time-dependent Schr\"odinger equation $i\hbar \frac{d\hat{U}(t, t_0)}{dt} = \hat{H}(t) \hat{U}(t, t_0)$ with initial condition $\hat{U}(t_0, t_0)=\hat{1}$. Here,  $\hat{U}_{\textrm{d}}(t,t_0)$  is the evolution operator of the laser-dressed system satisfying $i\hbar \frac{d\hat{U}_{\text{d}}(t, t_0)}{dt} = \hat{H}_{\text{LD}}(t) \hat{U}_{\text{d}}(t, t_0)$ with $\hat{U}_\text{d}(t_0, t_0)=\hat{1}$. In turn, $\hat{U}_{\textrm{p,I}}(t,t_{0})$ captures any additional contributions to the dynamics due to the probe laser in the presence of the drive. The decomposition becomes exact when $\hat{U}_{\textrm{p,I}}(t,t_{0})$ satisfies the time-dependent Schr\"odinger equation $i\hbar\frac{d}{dt}\hat{U}_{\textrm{p,I}}(t,t_0)=\hat{H}_{\textrm{p,I}}(t)\hat{U}_{\textrm{p,I}}(t,t_0)$ with $\hat{U}_{\textrm{p,I}}(t_0,t_0)=\hat{1}$, where $ \hat{H}_{\textrm{p,I}}(t)=\hat{U}_{\textrm{d}}^{\dagger}(t,t_0)\hat{H}_{\textrm{p}}(t)\hat{U}_{\textrm{d}}(t,t_0)$
is the interaction with the probe light in the interaction picture of  $\hat{H}_{\textrm{LD}}(t)$. This can be verified, for instance, by direct substitution into the time-dependent Schr\"odinger equation for $\hat{U}(t, t_0)$.

To understand the physical processes that contribute to $W(t,\omega)$ we introduce a transition amplitude between two many-body states $|\Psi_a\rangle$ and $|\Psi_b\rangle$ given by
\begin{equation}
\label{transamp}
A_{ba}=\left\langle \Psi_b \left| \hat{U}_{\textrm{d}}^{\dagger}\left(t, t_{0}\right)\hat{U}\left(t, t_{0}\right)\right| \Psi_a\right\rangle= \left\langle \Psi_{b} \left| \hat{U}_{\textrm{p,I}}(t,t_{0}) \right| \Psi_{a} \right\rangle .
\end{equation}
Equation \eqref{transamp} can be interpreted in two complementary but equivalent ways. It can be viewed as the overlap of the state of the system at time $t$ driven by both the drive and probe laser [i.e., $\hat{U}(t,t_0) | \Psi_{a} \rangle $] onto the laser-dressed state $\hat{U}_{\textrm{d}}(t,t_0) | \Psi_{b} \rangle$. Alternatively, it can be viewed as the projection onto state $| \Psi_{b} \rangle$ of an initial state $ |\Psi_{a} \rangle $ propagated forward in time $(t_{0} \rightarrow t)$ with both drive and probe laser turned on and then backward in time $(t \rightarrow t_{0})$ with only the drive laser on. 

We consider the effect of the probe light to first order in time-dependent perturbation theory. Thus, $\hat{U}_{\textrm{p,I}}(t,t_0)=\hat{1}+(\frac{-i}{\hbar})\int_{t_{0}}^{t} d t_{1} \hat{H}_{\textrm{p,I}}(t_{1})$ and
\begin{align}
A_{ba}&=\left\langle \Psi_b\left|\left(1-\frac{i}{\hbar} \int_{t_{0}}^{t} \hat{H}_{\textrm{p,I}}\left(t_{1}\right) d t_{1}\right)\right| \Psi_a\right\rangle .
\end{align}
Contributions to the transition probability $W(t,\omega)$ can arise due to transitions between different many-body states. That is,  
\begin{align}
\nonumber
W^{(1)}(t, \omega) &= \sum_{b\neq a}\left|A_{b a}\right|^{2} \\
& = \frac{1}{\hbar^{2}} \sum_{b\neq a}\left|\int_{t_{0}}^{t} d t_{1}\left\langle \Psi_b\left|\hat{H}_{\textrm{p,I}}\left(t_{1}\right)\right| \Psi_a\right\rangle\right|^{2},
\end{align}
where the sum runs over all many body states $|\Psi_b\rangle$ orthogonal to $|\Psi_a \rangle$ such that
\begin{align}
\label{comprelation}
& \sum_{b\neq a}|\Psi_b \rangle \langle \Psi_b |=\hat{1}-|\Psi_a \rangle \langle \Psi_a | .
\end{align}
A second contributing process to $W(t, \omega)$, $W^{(2)}(t, \omega)$, is due to the interaction of the probe laser with a permanent or induced dipole in the laser-dressed system leading to absorption and/or stimulated emission of a probe photon without changing the state in the laser-dressed material. This is, 
\begin{align}
\nonumber
W^{(2)}(t, \omega) &=\left|A_{aa}\right|^{2} \\
\nonumber
&=\left|1-\frac{i}{\hbar} \int_{t_{0}}^{t} d t_{1}\left\langle \Psi_a\left|\hat{H}_{\textrm{p,I}}\left(t_{1}\right)\right| \Psi_a\right\rangle\right|^{2} \\
\label{simpcomp}
&  =1+\frac{1}{\hbar^{2}} \left| \int_{t_{0}}^{t} d t_{1} \left\langle \Psi_a\left|\hat{H}_{\textrm{p,I}}\left(t_{1}\right)\right| \Psi_a\right\rangle\right|^{2} .
\end{align}

Combining the two processes and using the completeness relation Eq. \eqref{comprelation} yields the net probability that a probe photon is absorbed or emitted,
\begin{align}
\nonumber
&W(t, \omega)=W^{(1)}(t, \omega)+W^{(2)}(t, \omega) \\
\label{transprobinit}
&=\frac{1}{\hbar^{2}} \iint_{t_{0}}^{t} d t_{1} d t_{2}\left\langle \Psi_a\left|\hat{H}_{\textrm{p,I}}\left(t_{1}\right) \hat{H}_{\textrm{p,I}}\left(t_{2}\right)\right| \Psi_a\right\rangle+1 .
\end{align} 
The contribution from the constant term in Eq. \eqref{transprobinit} vanishes in $I(\omega$), thus we can drop it from this point on. Inserting Eq. \eqref{probeham} into Eq. \eqref{transprobinit} yields
\begin{align}
\nonumber
W(t,\omega)&=\frac{e^2 E_{\textrm{p}}^{2}}{\hbar^{2}m_{e}^{2}\omega^2} \iint_{t_{0}}^{t} dt_{1} dt_{2}  \left\langle \Psi_a \left| \hat{P}_{\textrm{I}}(t_1)\hat{P}_{\textrm{I}}(t_2) \right| \Psi_a\right\rangle \sin(\omega t_1)\sin(\omega t_2) \\
&=\frac{e^2  E_{\textrm{p}}^{2}}{2\hbar^{2}m_{e}^{2}\omega^2} \iint_{t_{0}}^{t} dt_{1} dt_{2} C_{P,P}(t_1,t_2)  \textrm{Re}[e^{-i\omega (t_1-t_2)}-e^{-i\omega( t_1+t_2)}] ,
\end{align}
where $\hat{P}_{\textrm{I}}(t)=\hat{U}_{\textrm{d}}^{\dagger}(t,t_0)\left(\hat{\mathbf{e}}_{\textrm{p}}\cdot\hat{\mathbf{P}}\right)\hat{U}_{\textrm{d}}(t,t_0) $,
and 
\begin{equation}
\label{twotimecoor}
C_{P,P}(t_1,t_2)=\left\langle \Psi_a \left| \hat{P}_{\textrm{I}}(t_1) \hat{P}_{\textrm{I}} (t_2) \right| \Psi_a \right\rangle 
\end{equation}
is the two-time momentum correlation function.
Therefore, the rate of transition is:
\begin{equation}
\label{rateofch}
I(\omega)=\lim_{t\rightarrow \infty}\frac{e^2 E_{\textrm{p}}^{2}}{2\hbar ^{2}m_{e}^{2}\omega^2(t-t_0)}\iint_{t_{0}}^{t} dt_{1} dt_{2} C_{P,P}(t_1,t_2) \textrm{Re}[e^{-i\omega (t_1-t_2)}-e^{-i\omega( t_1+t_2)}] .
\end{equation}
In the absence of drive laser (i.e. $E_{\textrm{d}}=0$), Eq. \eqref{rateofch} reduces to the well-know expression for the optical absorption $I_{\textrm{eq}}(\omega) \propto \int d\tau C(\tau)e^{-i\omega \tau}$ with $C(\tau)=\langle P(0)P(\tau)\rangle$ in linear response theory \cite{mukamel1995}. While for matter near equilibrium only the relative time is important, for laser-dressed matter the two-time correlation function ($C_{P,P}(t_1,t_2)$) is needed due to the breaking of time-translation symmetry by the driving laser. Equation \eqref{rateofch} can be solved numerically. However, that requires a computationally expensive two-time propagation. As discussed in Sec. \ref{sec2c}, a much more convenient approach is provided by Floquet theory.

To obtain the two-time correlation function [Eq. \eqref{twotimecoor}] we need to determine the dynamics of the momentum operator in interaction picture. In second quantization,
\begin{eqnarray}
\label{imoment}
\hat{P}_{\textrm{I}}(t) = \sum_{\mathbf{k}} \sum_{u, v} \langle \psi_{u \mathbf{k}}|\hat{\mathbf{e}}_{\textrm{p}}\cdot\hat{\mathbf{p}}|\psi_{v\mathbf{k}}\rangle \hat{c}^{\dagger}_{u\mathbf{k} , \textrm{I}} (t) \hat{c}_{v\mathbf{k},\textrm{I}}(t) ,
\end{eqnarray}
where
 \begin{equation}
 \label{annhil}
   \hat{c}_{u\mathbf{k},\textrm{I}}(t)=\sum_{v} \left( \hat{\mathcal{U}}(t,t_{0})\right)_{u\mathbf{k} , v\mathbf{k}} \hat{c}_{v\mathbf{k}} .
 \end{equation}
Here, $\mathcal{\hat{U}} (t,t_{0}) = \mathcal{T} e^{\frac{-i}{\hbar} \int_{t_0}^{t} \mathcal{\hat{H}}_{\textrm{LD}} (\tau) d\tau} $ is the time-ordered ($\mathcal{T})$ single-particle evolution operator which satisfies
\begin{equation}
     i\hbar\frac{d \hat{\mathcal{U}}(t,t_{0}) }{dt}=\mathcal{\hat{H}}_{\textrm{LD}}(t)\hat{\mathcal{U}}(t,t_{0}) 
\end{equation} 
with initial condition $\hat{\mathcal{U}}(t_{0},t_{0})=\hat{1} $. 
Substituting Eq. \eqref{annhil} into \eqref{imoment} and rearranging  we get
\begin{align}
 \label{epoper}
\hat{P}_{\textrm{I}}(t)&= \sum_{\mathbf{k}}\sum_{u,v,r,s}\langle\psi_{r\mathbf{k}}|\mathcal{\hat{U}}^{\dagger}(t,t_{0})|\psi_{u\mathbf{k}}\rangle \langle \psi_{u\mathbf{k}}|\hat{\mathbf{e}}_{\textrm{p}}\cdot \hat{\mathbf{p}}|\psi_{v\mathbf{k}}\rangle  \langle\psi_{v\mathbf{k}}|\mathcal{U}(t,t_{0})|\psi_{s\mathbf{k}}\rangle  \hat{c}^{\dagger}_{r\mathbf{k}} \hat{c}_{s\mathbf{k}} .
\end{align}
The problem of determining $\hat{P}_{\textrm{I}}(t)$ in Eq. \eqref{rateofch} via \eqref{epoper} is thus reduced to the problem of determining $\hat{\mathcal{U}} (t,t_{0}) $. Below we address this problem by using Floquet theory.

\subsection{Floquet Considerations}\label{sec2c}

Equation \eqref{rateofch}  defines the optical response of laser-dressed solids. However, numerically solving this equation is challenging because it requires propagating the many-body system to long times and back for several frequencies of the probe laser. To overcome this challenge, we now invoke the Floquet theory.

The single particle time-dependent Schr\"odinger equation for the laser-dressed solid is
\begin{eqnarray}
\label{oneelcsch}
i\hbar \frac{d}{d t}|\Psi (t) \rangle=\mathcal{\hat{H}}_{\textrm{LD}}(t)|\Psi (t)\rangle .
\end{eqnarray}
Since the Hamiltonian is periodic in both space $\mathcal{\hat{H}}_{\textrm{LD}} (\hat{\mathbf{r}} , t ) = \mathcal{\hat{H}}_{\textrm{LD}} ( \hat{\mathbf{r}} + \mathbf{R} , t ) $ and time  $\mathcal{\hat{H}}_{\textrm{LD}}(\hat{\mathbf{r}},t)=\mathcal{\hat{H}}_{\textrm{LD}}(\hat{\mathbf{r}},t+T)$, the system satisfies both Floquet \cite{Floquet1883} and Bloch theorem \cite{Ashcroft76}. Thus, the Floquet-Bloch states \cite{Faisal1997, Hsu2006, GomezLeon2013, Ikeda2018}
\begin{eqnarray}
\label{fbstate}
|\Psi_{\alpha \mathbf{k}}(t)\rangle=\frac{1}{\sqrt{V}} e^{-iE_{\alpha  \mathbf{k}} t/ \hbar} e^{ i\mathbf{k} \cdot \hat{\mathbf{r}} } | \Phi_{\alpha \mathbf{k}} (t) \rangle 
\end{eqnarray}
are solutions to the time-dependent Schr\"odinger Eq. \eqref{oneelcsch}. Here the  Floquet-Bloch mode $|\Phi_{\alpha \mathbf{k}}(t) \rangle$ with index $\alpha$ and crystal momentum $\mathbf{k}$ is a function that is periodic in both time and space  $\big( \Phi_{\alpha \mathbf{k}}(\mathbf{r},t+T)=\Phi_{\alpha \mathbf{k}}(\mathbf{r+R},t) =\Phi_{\alpha \mathbf{k}}(\mathbf{r},t) \big)$ and $E_{\alpha \mathbf{k}}$ is the corresponding quasienergy.

The Floquet-Bloch modes and quasienergies are determined by solving the following eigenvalue relation in Sambe space (tensor product of the regular Hilbert space and the space spanned by all $T$-periodic functions with basis $\{ e^{in\Omega t} \}$ where $n \in \mathbb{Z}$)
\begin{eqnarray}
\label{flqham}
\mathcal{\hat{H}}_{F} (\mathbf{k},\hat{\mathbf{r}},t) | \Phi_{\alpha k} (t) \rangle = E_{\alpha k} | \Phi_{\alpha k} (t) \rangle .
 \end{eqnarray}
Here, the Floquet-Bloch Hamiltonian is 
 \begin{align}
 \nonumber
   \mathcal{\hat{H}}_{F}(\mathbf{k},\hat{\mathbf{r}},t) & =\left[ e^{-i\mathbf{k} \cdot \hat{\mathbf{r}}} \mathbf{\mathcal{\hat{H}}}_{\textrm{LD}}(t) e^{i\mathbf{k} \cdot \hat{\mathbf{r}}} - i\hbar \frac{d}{d t} \right] \\
   \label{flqhamdefine}
   & = \frac{(\hat{\mathbf{p}} + \hbar \mathbf{k})^{2}}{2m_{e}} + V_{0} (\hat{\mathbf{r}})-\frac{eE_{\textrm{d}}}{m_{e}\Omega} \sin(\Omega t) \hat{\mathbf{e}}_{\textrm{d}} \cdot (\hat{\mathbf{p}} + \hbar \mathbf{k}) - i \hbar \frac{d}{d t} .
 \end{align}
 Equation \eqref{flqham} can be verified by substituting Eq. \eqref{fbstate} into \eqref{oneelcsch}.
 The Floquet-Bloch modes are uniquely defined in a Floquet-Brillouin zone (FBZ), for instance the fundamental FBZ being $\frac{-\hbar \Omega}{2}<E_{\alpha \mathbf{k}}\leq\frac{\hbar \Omega}{2}$. All other physically equivalent Floquet-Bloch states can be generated from the Floquet-Bloch modes and quasienergies within the fundamental FBZ \cite{Holthaus2015}.

Because of the time and space periodicity of the Floquet-Bloch modes, we can expand them in terms of their time Fourier components and the complete set of Bloch states
\begin{equation}
\label{ansatz2}
|\Phi_{\alpha \mathbf{k}}(t) \rangle = \sum_{n=-\infty }^{\infty} \sum_{u} F_{\alpha \mathbf{k}}^{(nu)} e^{in\Omega t} | u\mathbf{k} \rangle .
 \end{equation}
 The Bloch states $\{|u\mathbf{k}\rangle\}$ are eigenstates of the time-independent single-particle Hamiltonian
 \begin{equation}
\left[\frac{1}{2m_{e}}(\hat{\mathbf{p}}+\hbar \mathbf{k})^2+V_{0} (\hat{\mathbf{r}})\right]|u\mathbf{k}\rangle=\epsilon_{\mathbf{uk}}|u\mathbf{k}\rangle  .
\end{equation}
Substituting Eq. \eqref{ansatz2} into \eqref{flqham}, and using Eq. \eqref{flqhamdefine} we get 
\begin{eqnarray}
\nonumber
\sum_{n,u} F_{\alpha \mathbf{k}}^{(nu)} e^{in\Omega t} \left[ \frac{(\hat{\mathbf{p}} + \hbar \mathbf{k})^2}{2m_{e}} + V(\hat{\mathbf{r}}) - \frac{eE_{\textrm{d}}}{m_{e}\Omega} \sin(\Omega t) \hat{\mathbf{e}}_{\textrm{d}} \cdot  (\hat{\mathbf{p}} + \hbar \mathbf{k})  + n \hbar \Omega \right] | u\mathbf{k} \rangle \\
\label{flqhameig}
  = E_{\alpha \mathbf{k}}\sum_{n,u}F_{\alpha \mathbf{k}}^{(nu)}e^{in\Omega t}|u\mathbf{k}\rangle .
\end{eqnarray}
Left multiplying by $\langle v\mathbf{k}|e^{-im\Omega t} $ and integrating over one time period $T=\frac{2\pi} {\Omega}$ $\left( \textrm{that is, } \frac{1}{T}\int_{0}^{T} dt \cdots \right)$ yields the eigenvalue equation
\begin{equation}
\label{flqeig}
\sum_{n,u}\Gamma_{nu,mv,\mathbf{k}}F_{\alpha \mathbf{k}}^{(nu)}=E_{\alpha \mathbf{k}}F_{\alpha \mathbf{k}}^{(mv)} ,
\end{equation}
where
\begin{eqnarray}
\label{finalflq}
\Gamma_{nu,mv,\mathbf{k}}= (\epsilon_{u\mathbf{k}}+n\hbar\Omega )\delta_{nm}\delta_{uv} -\frac{e E_{\textrm{d}}}{2 i m_{e}\Omega} \hat{\mathbf{e}}_{\textrm{d}} \cdot \mathbf{p}_{u\mathbf{k},v\mathbf{k}} (\delta_{n,m-1}-\delta_{n,m+1}).
\end{eqnarray}
For a given $\mathbf{k}$, Eq. \eqref{flqeig} defines an eigenvalue problem that yields the quasienergies as eigenvalues and the Floquet-Bloch modes as eigenvectors. The Floquet-Bloch states are obtained from the Floquet-Bloch modes using Eq. \eqref{fbstate}.

These quantities define the single-particle evolution operator \cite{Shirley1965}
\begin{eqnarray}
\label{evolut}
\mathcal{\hat{U}}(t,t_{0})=\sum_{\mathbf{k},\alpha}|\Psi_{\alpha \mathbf{k}}(t)\rangle \langle \Psi_{\alpha \mathbf{k}}(t_0)| 
\end{eqnarray}
needed to calculate the two-time correlation function in Eq. \eqref{epoper}, as detailed below.

\subsection{Computing the Two-time Correlation Function}\label{sec2d}

Substituting Eq. \eqref{evolut} into \eqref{epoper} we get
\begin{align}
\nonumber
 \hat{P}_{\textrm{I}}(t) &= \sum_{\mathbf{k,k',k''}}\sum_{u,v,r,s}\sum_{\alpha, \beta} \langle \psi_{r\mathbf{k}} | \Psi_{\alpha \mathbf{k'}}(t_0) \rangle \langle \Psi_{\alpha \mathbf{k'}}(t) | \psi_{u\mathbf{k}} \rangle \\
 \nonumber
  & \times \langle \psi_{u\mathbf{k}} | \hat{\mathbf{e}}_{\textrm{p}} \cdot \hat{\mathbf{p}} | \psi_{v\mathbf{k}} \rangle \langle \psi_{v \mathbf{k}}  | \Psi_{\beta \mathbf{k''}}(t) \rangle \langle \Psi_{\beta \mathbf{k''}}(t_0)  | \psi_{s\mathbf{k}} \rangle  \hat{c}^{\dagger}_{r\mathbf{k}} \hat{c}_{s\mathbf{k}} \\
\label{epoperlast}
& = \frac{1}{V^{2}}
\sum_{\mathbf{k}}\sum_{u,v}\sum_{\alpha ,\beta}e^{iE_{\alpha \beta \mathbf{k}}(t-t_0)/\hbar}\langle u\mathbf{k}|\Phi_{\alpha \mathbf{k}}(t_{0})\rangle \langle\Phi_{\beta \mathbf{k}}(t_{0})| v\mathbf{k}\rangle \mathcal{P}_{\alpha\beta \mathbf{k}}(t)  \hat{c}^{\dagger}_{u\mathbf{k}} \hat{c}_{v\mathbf{k}} ,
\end{align}
where $E_{\alpha \beta \mathbf{k}}=E_{\alpha \mathbf{k}}-E_{\beta  \mathbf{k}}$, and where we have taken into account Eqs. \eqref{fbstate} and \eqref{ansatz2}, and the orthonormality of Bloch states $\langle \psi_{u\mathbf{k}} | \psi_{v\mathbf{k'}} \rangle = \delta_{u v}\delta_{\mathbf{k} \mathbf{k'}}$. Here we define the momentum matrix elements (MME) between the Floquet-Bloch modes $\alpha, \beta$ with crystal momentum $\mathbf{k}$ as
\begin{eqnarray}
\mathcal{P}_{\alpha\beta \mathbf{k}}(t)=\frac{1}{V} \langle\Phi_{\alpha \mathbf{k}}(t)|\hat{\mathbf{e}}_{\textrm{p}}\cdot (\hat{\mathbf{p}}+\hbar \mathbf{k})|\Phi_{\beta \mathbf{k}}(t)\rangle  .
\end{eqnarray}

The Floquet-Bloch modes and their MME are $T$-periodic. Hence, we can expand them in a Fourier series given by
\begin{eqnarray}
\label{momperiodic}
\mathcal{P}_{\alpha \beta \mathbf{k}}(t)=\sum_{n=-\infty}^{\infty}\mathcal{P}_{\alpha\beta \mathbf{k}}^{(n)}e^{in\Omega t} ,
\end{eqnarray}
where 
\begin{equation}
    \label{fbmme}
\mathcal{P}_{\alpha\beta \mathbf{k}}^{(n)}=\frac{1}{T}\int_{0}^{T}dt \mathcal{P}_{\alpha\beta \mathbf{k}}(t)e^{-in\Omega t} 
\end{equation}
is the $n$th Fourier component. Substituting Eq. \eqref{momperiodic} into \eqref{epoperlast} gives
\begin{align}
\label{imomentfinal}
\hat{P}_{\textrm{I}}(t)&= \frac{1}{V^{2}} \sum_{\mathbf{k}}\sum_{u,v}\sum_{\alpha ,\beta}\sum_{n}e^{iE_{\alpha\beta \mathbf{k}}(t-t_{0})/\hbar +in\Omega t}D_{uv, \alpha \beta , \mathbf{k}}^{(n)} \hat{c}^{\dagger}_{u\mathbf{k}} \hat{c}_{v\mathbf{k}} ,
\end{align}
where $ D_{u v,\alpha \beta ,\mathbf{k}}^{(n)} =\langle u\mathbf{k}|\Phi_{\alpha \mathbf{k}}(t_{0})\rangle \langle\Phi_{\beta \mathbf{k}}(t_{0})| v\mathbf{k}\rangle\mathcal{P}_{\alpha\beta \mathbf{k}}^{(n)} $ .

Using Eq. \eqref{imomentfinal} we can now obtain the two-time momentum correlation function Eq. \eqref{twotimecoor} 
\begin{align}
C_{P,P}(t_1,t_2) & = \left\langle \Psi_{a} \left| \hat{P}_{\textrm{I}}(t_{1})\hat{P}_{\textrm{I}}(t_{2}) \right|  \Psi_{a} \right\rangle \\
\nonumber
& =\frac{1}{V^{4}} \sum_{\mathbf{k}}\sum_{u,v,\alpha,\beta}\sum_{u',v',\gamma,\delta}\sum_{n,m} D_{uv,\alpha\beta, \mathbf{k}}^{(n)}D_{u'v',\gamma \delta ,\mathbf{k}}^{(m)} \\
\label{cpp}
& \times e^{i\left( \frac{E_{\alpha\beta \mathbf{k}}}{\hbar} (t_1-t_{0}) + n\Omega t_1\right)}  e^{ i \left( \frac{E_{\gamma\delta \mathbf{k}}}{\hbar} (t_2-t_{0}) +  m \Omega t_2 \right) }  \langle \Psi_{a} | 
 \hat{c}_{u\mathbf{k}}^{\dagger} \hat{c}_{v\mathbf{k}} \hat{c}_{u'\mathbf{k}}^{\dagger} \hat{c}_{v'\mathbf{k}} | \Psi_{a} \rangle  .
\end{align}
The initial occupation factor
\begin{eqnarray}
\label{capitaln}
\langle \Psi_{a}  |\hat{c}_{u\mathbf{k}}^{\dagger}\hat{c}_{v\mathbf{k}}\hat{c}_{u'\mathbf{k}}^{\dagger}\hat{c}_{v'\mathbf{k}}| \Psi_{a} \rangle =\delta_{u v} \delta_{u' v'} \bar{n}_{u \mathbf{k}}  \bar{n}_{u' \mathbf{k}} + \delta_{u v'} \delta_{u' v} \bar{n}_{u \mathbf{k}} (1 - \bar{n}_{u' \mathbf{k}}) = N_{u,v,u',v',\mathbf{k}} ,
\end{eqnarray}
where $\bar{n}_{u \mathbf{k}}=\langle \Psi_{a} | \hat{c}_{u \mathbf{k}}^{\dagger}\hat{c}_{u \mathbf{k}} | \Psi_{a} \rangle $ is the initial particle occupation in band $u$ at crystal momentum $\mathbf{k}$ as determined by the Fermi-Dirac distribution.

\subsection{Optical Absorption Formula}\label{sec2e}

The rate of transition induced due to a probe photon in the laser-dressed system is obtained by substituting Eq. \eqref{cpp} into \eqref{rateofch} to yield

\begin{eqnarray}
\nonumber
I(\omega)=\lim_{t\rightarrow \infty}\frac{e^2E_{\textrm{p}}^{2} }{2 V^{4}\hbar^{2}m_{e}^{2}\omega^2} \frac{1}{(t-t_{0})}\iint_{t_{0}}^{t}dt_{1} dt_{2} \sum_{\mathbf{k}}\sum_{u,v,\alpha,\beta}\sum_{u',v',\gamma,\delta}\sum_{n,m} D_{uv,\alpha\beta, \mathbf{k}}^{(n)}D_{u'v',\gamma \delta, \mathbf{k}}^{(m)} \\
\times e^{i \left( \frac{E_{\alpha\beta \mathbf{k}}}{\hbar}(t_1-t_{0})+ n \Omega t_1 \right) }e^{i \left( \frac{E_{\gamma\delta \mathbf{k}}}{\hbar} (t_2-t_{0}) + m \Omega t_2 \right) }  N_{u,v,u',v',\mathbf{k}}  \textrm{Re} [e^{-i\omega (t_1-t_2)}-e^{-i\omega( t_1+t_2)}] .
\end{eqnarray}
To further simplify the above double time integral, we transform it in terms of a center of mass time $\bar{t}=\frac{t_{1} + t_{2} }{2}$ and a relative time $\tau = t_{2}-t_{1}$. We also take the preparation time of the system to be in remote past, such that $t_{0} \rightarrow -\infty$. This gives
\begin{eqnarray}
\nonumber
I(\omega)=\lim_{t\rightarrow \infty} \frac{e^2E_{\textrm{p}}^{2} }{2V^{4} \hbar^{2}m_{e}^{2}\omega^2} 
 \frac{1}{(t-t_{0})} \iint_{-\infty}^{t}d\bar{t}d\tau \sum_{\mathbf{k} } \sum_{u,v,\alpha,\beta} \sum_{u',v',\gamma,\delta} \sum_{n,m} D_{uv,\alpha\beta ,\mathbf{k}}^{(n)} D_{u'v',\gamma \delta ,\mathbf{k}}^{(m)} \\ 
\label{netabsbothterm}
\times e^{i[(\frac{E_{\gamma\delta \mathbf{k}}}{\hbar}+\frac{E_{\alpha\beta \mathbf{k}}}{\hbar})(\bar{t}-t_{0})]+i(m+n)\Omega \bar{t}}e^{i[(\frac{E_{\gamma\delta \mathbf{k}}}{\hbar}-\frac{E_{\alpha\beta \mathbf{k}}}{\hbar})+(m-n)\Omega] \tau /2}  N_{u,v,u',v',\mathbf{k}}  \textrm{Re} [e^{i\omega \tau}-e^{2i\omega \bar{t}}] .
\end{eqnarray}
The terms proportional to $e^{2i\omega \bar{t}}$ above will generally not contribute to $I(\omega)$ (Appendix in Ref. \cite{Gu2018}), thus we get
\begin{eqnarray}
\nonumber
I(\omega)=\lim_{t\rightarrow \infty} \frac{e^2E_{\textrm{p}}^{2} }{4 V^{4} \hbar^{2}m_{e}^{2}\omega^2}  \frac{1}{(t-t_{0})}  \iint_{-\infty}^{t} d\bar{t} d\tau  \sum_{\mathbf{k} } 
 \sum_{u,v,\alpha,\beta} \sum_{u',v',\gamma,\delta} \sum_{n,m} D_{uv,\alpha\beta ,\mathbf{k}}^{(n)} D_{u'v',\gamma \delta,\mathbf{k}}^{(m)} \\ 
\times e^{i[(\frac{E_{\gamma\delta \mathbf{k}}}{\hbar}+\frac{E_{\alpha\beta \mathbf{k}}}{\hbar})(\bar{t}-t_{0})]+i(m+n)\Omega \bar{t}}e^{i[(\frac{E_{\gamma\delta \mathbf{k}}}{\hbar}-\frac{E_{\alpha\beta \mathbf{k}}}{\hbar})+(m-n)\Omega] \tau /2}  N_{u,v,u',v',\mathbf{k}} (e^{i\omega \tau}+e^{-i\omega \tau}) . 
\end{eqnarray}
The integral over $\bar{t}$ as $t \rightarrow \infty $ is zero due to the oscillatory terms and the $1/(t-t_{0})$ factor except when $E_{\gamma\delta \mathbf{k}} + E_{\alpha\beta \mathbf{k}}=0$ and $m+n=0$. For this,  $n=-m$ and the Floquet-Bloch modes are such that either $\delta = \gamma$ and $\beta = \alpha$ or $\beta = \gamma $ and $\delta = \alpha$. Using these conditions we get
\begin{align}
\nonumber
I(\omega) & =\frac{e^2E_{\textrm{p}}^{2} }{4 V^{4} \hbar^{2}m_{e}^{2}\omega^2}   \int_{-\infty}^{\infty} d\tau  \sum_{\mathbf{k}} \sum_{u,v,u',v'} \sum_{\alpha,\gamma} \sum_{m} \left( D_{uv,\alpha\alpha, \mathbf{k}}^{(-m)} D_{u'v',\gamma \gamma ,\mathbf{k}}^{(m)} e^{im\Omega \tau }  \right.  \\
&  \left. +   D_{uv,\alpha\gamma ,\mathbf{k}}^{(-m)}D_{u'v',\gamma \alpha ,\mathbf{k}}^{(m)} 
 e^{i[\frac{E_{\gamma \alpha \mathbf{k}}}{\hbar}+m\Omega] \tau } \right) N_{u,v,u',v',\mathbf{k}} (e^{i\omega \tau}+e^{-i\omega \tau})   .
\end{align}
Performing the integral with respect to $\tau$ and relabeling dummy variables
\begin{eqnarray}
\nonumber
\label{prefinal}
I(\omega)=\frac{e^2 E_{\textrm{p}}^{2}\pi}{2 V^{4} \hbar m_{e}^{2}\omega^2} \sum_{\mathbf{k}}\sum_{u, v, u', v'}\sum_{\alpha, \beta} \sum_{n} \left( D^{(-n)}_{uv,\alpha \alpha ,  \mathbf{k}}D^{(n)}_{u'v',\beta\beta , \mathbf{k}}\delta(n\hbar\Omega-\hbar \omega) \right. \\
\left. + D^{(-n)}_{uv ,\beta \alpha , \mathbf{k}}D^{(n)}_{u'v',\alpha\beta, \mathbf{k}}\delta(E_{\alpha\beta \mathbf{k}}+n\hbar\Omega-\hbar \omega) \right) N_{u,v,u',v',\mathbf{k}}+(\omega\leftrightarrow -\omega) ,
\end{eqnarray}
where $(\omega \leftrightarrow -\omega)$ represents terms that are equal to the first two terms but with  $\omega$ replaced with $-\omega$. To obtain Eq. \eqref{prefinal} we have used the properties of the Dirac delta function $\int_{-\infty}^{\infty}  e^{i\omega t} dt = 2\pi \delta(\omega)$  and  $ \delta(x) = \hbar \delta(\hbar x)$. 

Equation \eqref{prefinal} captures the transition rate from all  processes occurring in the laser-dressed system due to interaction with the probe photon. However, it does not distinguish  absorption from stimulated emission. We identify the first two terms with ($-\hbar \omega$) in the delta functions as terms leading to absorption and the remaining terms with ($+\hbar \omega$) in the delta functions representing stimulated emission. Therefore, the  net absorption rate $R(\omega)$ is defined as the difference between rate of absorption and stimulated emission. That is,
\begin{align}
\nonumber
& R(\omega)=\frac{e^2 E_{\textrm{p}}^{2}\pi}{2 V^{4} \hbar m_{e}^{2}\omega^2} 
 \sum_{\mathbf{k}} \sum_{u,u'} \sum_{\alpha, \beta} \sum_{n} D^{(-n)}_{ uu' , \beta \alpha , \mathbf{k}} D^{(n)}_{ u'u , \alpha\beta,\mathbf{k}} \\
& \times [\delta(E_{\alpha \beta \mathbf{k}}+n\hbar\Omega-\hbar \omega)-\delta(E_{\alpha\beta \mathbf{k}}+n\hbar\Omega+\hbar \omega)]\bar{n}_{u}(1-\bar{n}_{u'}) ,
\end{align}
where the contribution coming from the first term in Eq. \eqref{prefinal} and those involving $\bar{n}_{u\mathbf{k}}\bar{n}_{u'\mathbf{k}}$ exactly cancel. We define the quantity
\begin{eqnarray}
\label{finalpopfact}
\Lambda_{ \alpha \beta \mathbf{k}}=\frac{1}{V^{4}}\sum_{u', u}|\langle u\mathbf{k}|\Phi_{\beta \mathbf{k}}(t_{0})\rangle |^2 |\langle\Phi_{\alpha \mathbf{k}}(t_{0})|u'\mathbf{k}\rangle |^2 \bar{n}_{u \mathbf{k}}(1-\bar{n}_{u' \mathbf{k}}) 
\end{eqnarray}
as the population factor for Floquet-Bloch $\beta \rightarrow \alpha$ transition at crystal momentum $\mathbf{k}$ . Since the harmonics of the MME satisfy $\mathcal{P}_{\alpha\beta \mathbf{k}}^{(n)}=\mathcal{P}_{\beta\alpha\mathbf{k}}^{(-n)*}$, then
\begin{align}
\label{netrate}
& R(\omega)=\frac{e^2E_{\textrm{p}}^{2} \pi}{2\hbar m_{e}^{2}\omega^2} \sum_{\mathbf{k}} \sum_{\alpha, \beta}\sum_{n}\Lambda_{\alpha \beta \mathbf{k}}|\mathcal{P}_{\alpha\beta \mathbf{k}}^{(n)}|^2  
[\delta(E_{\alpha\beta \mathbf{k}}+n\hbar\Omega-\hbar \omega)-\delta(E_{\alpha\beta \mathbf{k}}+n\hbar\Omega+\hbar \omega)] .
\end{align}

The optical absorption coefficient $A(\omega)$ is defined as $A(\omega) = \frac{R(\omega) \hbar \omega }{V I_{0}} $. That is, as the ratio of power absorbed from the incident probe laser per unit volume $V$ and  incident light flux $I_{0}=\epsilon_{0}E_{\textrm{p}}^2c n_{r}/2$, where $\epsilon_{0}$ is the permittivity of vacuum, $c$ the speed of light and $n_{r}$ is the refractive index of the material \cite{Dresselhaus2018}.
From Eq. \eqref{netrate}, we can thus write
\begin{align}
    \label{final}
    A(\omega)    & =  \frac{e^{2}\pi}{ m_{e}^2 \epsilon_{0}c n_{r} V \omega}\sum_{\mathbf{k}}\sum_{\alpha, \beta}\sum_{n}\Lambda_{\alpha \beta \mathbf{k}}|\mathcal{P}_{\alpha\beta \mathbf{k}}^{(n)}|^2  
  [\delta(E_{\alpha\beta \mathbf{k}}+n\hbar\Omega-\hbar \omega)-\delta(E_{\alpha\beta \mathbf{k}}+n\hbar\Omega+\hbar \omega)].
\end{align}

Equation \eqref{final} defines the linear optical absorption of laser-dressed solids and is the main result of this paper. An absorption or stimulated emission event occurs when the probing photon energy $\hbar \omega$ coincides with a Bohr transition energy between two Floquet-Bloch modes $E_{\alpha \beta \mathbf{k}}+n\hbar\Omega$. The first term in Eq. \eqref{final} leads to absorption while the second term captures stimulated emission. The intensity of a transition from $\beta \rightarrow \alpha$ Floquet-Bloch modes separated by $n$ Floquet-Brillouin zones is determined by the population factor $\Lambda_{\alpha \beta \mathbf{k}}$ and the Fourier components of the MME $\mathcal{P}_{\alpha\beta \mathbf{k}}^{(n)}$. The population factor captures population changes due to the drive and  guarantees that an optical transition happens only from an initially occupied Floquet-Bloch mode to an empty one. In turn, the MME determines the strength of the transition and it depends not only on the states involved but also on the number $n$ of FBZ that separate the two states. Equation \eqref{final} also shows that the optical absorption of laser-dressed solid is not solely determined by the density of states of the driven system as the momentum matrix elements and population factors also play a pivotal role in determining $A(\omega)$.

It is instructive to contrast Eq. \eqref{final} to the usual equilibrium  absorption coefficient for solids given by $ \alpha(\omega)\propto \sum_{\mathbf{k}} (f(E_{v\mathbf{k}})-f(E_{c\mathbf{k}})) |\langle v|\hat{\mathbf{p}}|c\rangle |^2 \delta({E_{c\mathbf{k}}-E_{v\mathbf{k}}-\hbar\omega})$ \cite{Dresselhaus2018}, where $v,c$ represent labels for valence and conduction band respectively, $E_{c/v\mathbf{k}}$ is the energy of the conduction/valence band and $f(E)$ is the Fermi-Dirac distribution. Equation \eqref{final} is reminiscent to the equilibrium case except that the Floquet-Bloch modes in the laser-dressed system play the role of pristine eigenstates. That is, optical transitions can be viewed as occurring between the Floquet-Bloch modes. There are three additional new features. First, the population factor $\Lambda_{\alpha \beta \mathbf{k}}$ depends on the drive. That is, the driving changes the set of states that are accessible for the probe laser. In addition the transition MME depend on the number $n$ of FBZ between the two modes involved. Last, Eq. \eqref{final} predicts the emergence of replicas of a given transition separated by integers of the drive photon energy $n\hbar\Omega$. This is because transitions in the laser-dressed system 
 can now occur among the Floquet-Bloch modes across different FBZ. Overall Eq. \eqref{final} shows that the Floquet-Bloch modes are the natural states to understand the non-equilibrium absorption properties of periodically driven solids.

\section{Computational approach}\label{sec3}

The theory in Sec. \ref{sec2} is general and can be used to compute the optical properties of laser-dressed solids with arbitrary band structure and dimensionality. As input, the theory uses the band structure of the solid and the MME between the Bloch modes which can be obtained from first-principle electronic structure calculations. To compute the optical absorption spectrum of a laser-dressed solid we have developed a FORTRAN based code named \emph{FloqticS} --Floquet optics in solids-- that is accessible through \texttt{GitHub} \cite{code}. 

The code requires the following input characterizing the system: the number of $\mathbf{k}$ vectors and their values in the Brillouin zone, number of bands ($N_{1}$) and their energies $\epsilon_{u\mathbf{k}}$, 
and the initial occupation numbers for the bands ($\bar{n}_{u \mathbf{k}}$).  To characterize the light-matter interaction, the code requires the  number of time-periodic functions used as a basis in the calculation ($N_{2}$), the MME among Bloch states in the direction of drive laser polarization $(\hat{\mathbf{e}}_{\textrm{d}} \cdot \mathbf{p}_{u\mathbf{k},v\mathbf{k}} )$, and in the direction of probe laser polarization $(\hat{\mathbf{e}} _{\textrm{p}}\cdot \mathbf{p}_{u\mathbf{k},v\mathbf{k}} )$, the drive laser photon energy $\hbar\Omega$, and its amplitude $E_{\textrm{d}}$. All $\epsilon_{u\mathbf{k}}$, $\bar{n}_{u \mathbf{k}}$, and MME should be ordered according to the $\mathbf{k}$ vectors. The MME among the Bloch states are defined using the definition of Bloch states in Sec. \ref{sec2}(a) and Eq. \eqref{momentumcoup}. For each $\mathbf{k}$, the dimension of Sambe space is equal to $N_{1}N_{2}$. The quantity $N_{2}$ is a parameter that needs to be increased until the Floquet-Bloch modes and quasienergies are converged. Additionally, the number of bands and the number of $\mathbf{k}$ vectors should also be varied to attain a converged absorption spectrum of the laser-dressed solid at a given drive laser parameters.
 
 With all these inputs, the code proceeds to perform the following for each provided $\mathbf{k}$ vector: It constructs the Floquet Hamiltonian in the Floquet-Bloch mode basis using  Eq. \eqref{finalflq} and diagonalizes it to obtain the coefficients $F_{\alpha \mathbf{k}}^{(nu)}$ and quasienergies $E_{\alpha \mathbf{k}}$. This leads to $N_{1}N_{2}$ Floquet-Bloch modes and quasienergies but the code only stores the physically relevant $N_{1}$ quasienergies in the fundamental FBZ ($\frac{-\hbar \Omega}{2}<E_{\alpha \mathbf{k}}\leq\frac{\hbar \Omega}{2}$) and corresponding Floquet-Bloch modes  for further computation. The code computes the population factor $\Lambda_{\alpha\beta \mathbf{k}}$ among all the modes using  Eqs. \eqref{ansatz2} and \eqref{finalpopfact} and the provided occupation numbers. The code also computes the Fourier components of MME among the Floquet-Bloch modes $\mathcal{P}_{\alpha\beta \mathbf{k}}^{(n)}$. There will be $2 N_{2}+1$ (that is, number of integers $\in [-N_{2},  N_{2}]$) number of Fourier components in this case. The code then calculates the absorption spectrum using Eq. \eqref{final} and reports the intensity of transition $A(\omega)$ as  a function of $\hbar\omega$ for each $\mathbf{k}$ vector. This whole procedure is repeated for all provided $\mathbf{k}$ vectors.

 The major bottleneck in the computation is the diagonalization of the Floquet Hamiltonian. For example, a well-converged  computation with nonresonant drive with $  N_{1} = 100$ bands requires $~ N_{2}=1000$ time-periodic functions for convergence. The dimensions of the Floquet Hamiltonian to be diagonalized  $Z^{2}=(N_{1}\times N_{2})^{2} = 10^{5} \times 10^{5}$. A usual diagonalization algorithm such as $\texttt{ZHEEV}$ in LAPACK  which scales with $\mathcal{O}(Z^{3})$  become computationally unfeasible. To solve this issue, we have incorporated the parallelized diagonalization package ELPA \cite{Marek2014} into \emph{FloqticS}.  A block-cyclic distribution of the Floquet Hamiltonian is employed as input to the ELPA package. This parallelizes the diagonalization in both time and memory as only a part of the whole Floquet matrix is stored into each node. \emph{FloqticS} only collects the eigenvectors and eigenvalues in a FBZ from all nodes and sends it to the root node to perform the final computation of the absorption coefficient. The efficient diagonalization through ELPA allows us to compute the absorption properties with a finer Brillouin zone sampling of a  realistic solid in a tractable computational time.

\section{Laser-dressed one-dimensional solid}\label{sec4}

\subsection{Hamiltonian Model and Computational Details}\label{sec4a}

To illustrate the theory and emerging physics, below we compute the optical absorption coefficient Eq. \eqref{final} for an exemplifying one-dimensional solid with Hamiltonian $\hat{H}_{0}=\frac{\hat{p}^2}{2m_{e}}+V(\hat{x})$. We consider a cosine-shaped lattice potential $V(\hat{x}) = V_{0} \left(1+\cos(\frac{2\pi \hat{x}}{a_{0}}) \right) $, where $a_{0}$ is the unit-cell length. 
 Such a model has been used thoroughly before  \cite{Wang2020, Hawkins2015, Ernotte2018, Wu2015, Lang2022, Yue2020} to study the properties of laser-driven solids. The advantage of this model with respect to tight-binding Hamiltonians is that it enables computation with an arbitrary number of bands as required to test convergence. For definitiveness, we take the potential depth to be $V_{0}=-10.06$ eV and $a_{0}=4.23$ \r{A} which yields a 4.18 eV band gap, which is representative of a wide band-gap semiconductor.

The Bloch states and band structure are determined by the time-independent Schr\"odinger equation
\begin{eqnarray}
\label{blocheigen}
\left[\frac{\hat{p}^2}{2m_{e}}+V_{0}\left(\hat{1}+\cos\left(\frac{2\pi \hat{x}}{a_{0}}\right)\right)\right]|\psi_{uk}\rangle=\epsilon_{uk}|\psi_{uk}\rangle .
\end{eqnarray}
 We diagonalize this Hamiltonian using the Bloch states  $|\psi_{uk}\rangle = \frac{1}{\sqrt{V}} e^{ik\hat{x}} |uk \rangle $ where the  Bloch functions $\langle x| uk \rangle=\langle x+a_{0}| u k \rangle = \sum_{K}c_{u,k-K}e^{-iK x} $ are expanded in a plane wave basis and $V=Ma_{0}$, where $M$ is the total number of unit cells that compose the supercell. Here, $K$ is the set of reciprocal-space lattice vectors given by integer multiples of $\frac{2\pi}{a_{0}}$. The number of vectors in the set $\{ K \}$ determine the number of bands to be obtained from Eq. \eqref{blocheigen}. The $k$ points in the first Brillouin zone are determined by the Born-Von Karman periodic boundary condition $k=\frac{2\pi j}{a_{0}M}$ for one dimension \cite{Ashcroft76} where $ j \textrm{ is integer} \in [-M/2,M/2)$. This leads to an eigenvalue problems for each $k$ point in the Brillouin zone, which provides the band structure $\epsilon_{uk}$ and coefficients $c_{u,k-K}$. We further compute the MME among the Bloch states using  
\begin{equation}
\label{blochmme}
\langle \psi_{uk}| \hat{p} | \psi_{vk}\rangle =\frac{1}{V} \langle uk| (\hat{p}+\hbar k ) | vk \rangle =  \sum_{K}c^{*}_{u,k-K}c_{v,k-K}(\hbar k-\hbar K) .
\end{equation}

\begin{figure}[htbp]
\begin{center}
\includegraphics[width=0.5\textwidth]{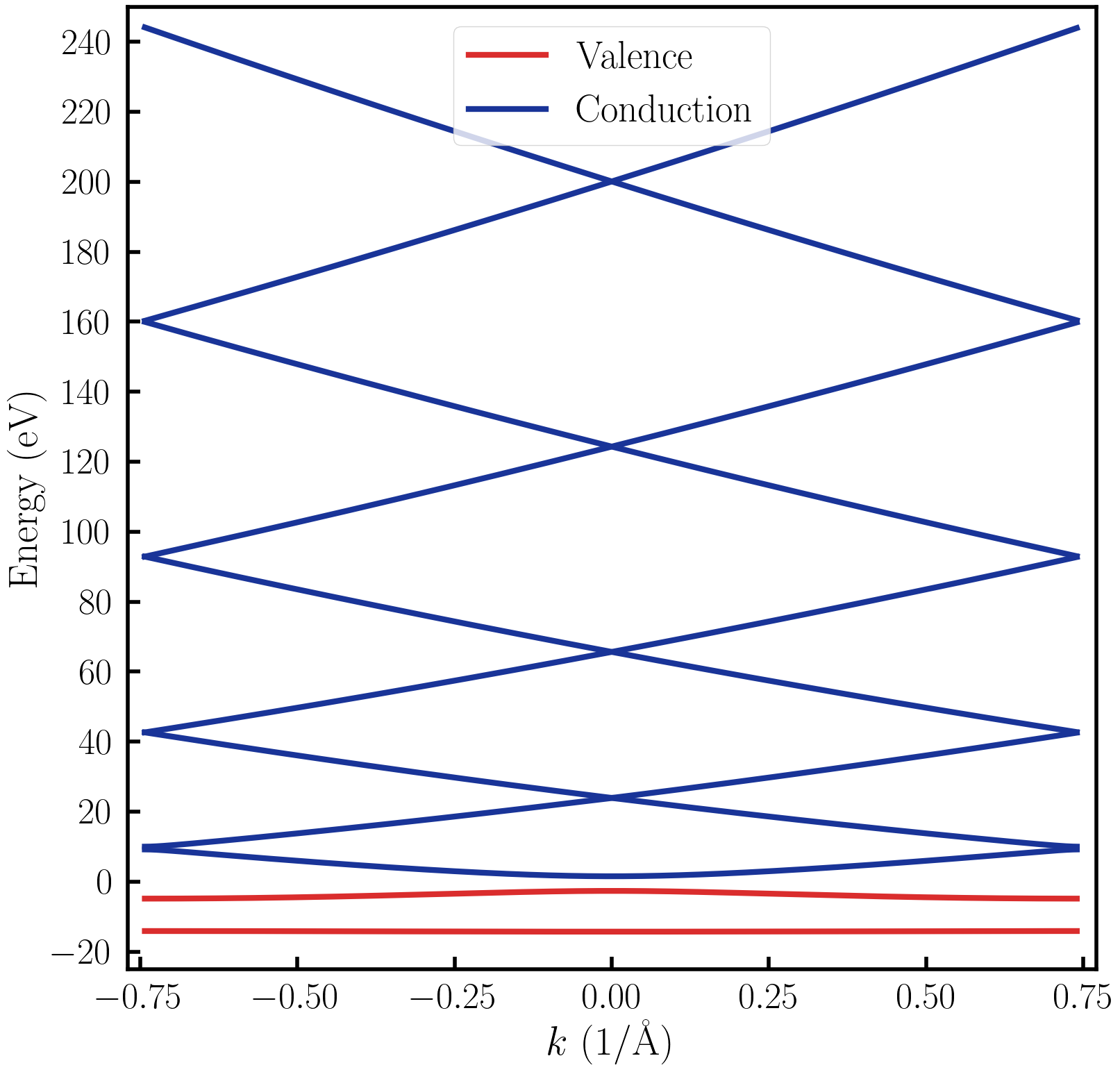}
\caption{ Band structure of cosine-shaped lattice potential in the first Brillouin zone showing the 11 bands taken into account in the calculation. The Fermi energy is taken at 0.0 eV and the valence bands are shown in red and conduction bands in blue. The direct band gap of  $4.18$ eV is located at the $\Gamma$ point $k=0$ \r{A}$^{-1}$}
\label{bandst}
\end{center}
\end{figure}

The eigenvalue problem Eq. \eqref{blocheigen} can provide all bands at each $k$ point.  In practice, one needs to truncate to a finite number of bands that provides converged results. For the electric field amplitude up to $0.4$ V/\r{A} and drive photon energy $0.5$ eV employed here, we find that 11 bands (two valence and nine conduction) and 1201 Floquet channels [-600 to 600 $n$ in Eq. \eqref{flqeig}] provide convergence.  Such convergence checks are particularly important for field-driven solids as Hilbert space truncation can lead to violation of gauge invariance \cite{Taghizadeh2017}. The band structure for the model is sampled by 500 $k$ points in the Brillouin zone shown in Fig. \ref{bandst}.  We define  the Fermi energy at 0.0 eV and obtain a direct band gap of 4.18 eV located at the $\Gamma$ point $(k=0$ \r{A}$^{-1})$.

The drive laser photon energy $\hbar \Omega=0.5$ eV is chosen to be non-resonant to suppress  near-resonant interband multiphoton absorption. In this way, the laser-dressing transiently distorts the electronic structure and the solid can reversibly return to its initial state by turning off the drive laser for the $E_{\textrm{d}}$ highlighted here. The drive and probe laser polarization direction is chosen to be along the crystal growth direction. 

The band structure $\epsilon_{uk}$ and MME Eq. \eqref{blochmme} along with the drive laser parameters provide all the information needed to solve the Floquet eigenvalue problem in Eq. \eqref{flqeig}, compute the population factor in Eq. \eqref{finalpopfact} and the MME among the Floquet-Bloch modes in Eq. \eqref{fbmme}.  These quantities are then used in Eq. \eqref{final} to compute the laser-dressed optical absorption spectra for each $k$ point in the first Brillouin zone. The 11 bands taken into account in the calculation lead to a total of 11 Floquet-Bloch modes for each $k$ point in a FBZ. The transition peaks with transition energy below 0.03 eV are removed from the calculation to obtain meaningful results in the low-frequency region as the results are limited by the smoothness of the $k$-grid used to sample  in the Brillouin zone. For simplicity we take $n_{r}=1$ to be independent of probe frequency.

\subsection{Results and Discussion}\label{4c}

\begin{figure*}[htbp]
\begin{center}
\includegraphics[width=\textwidth]{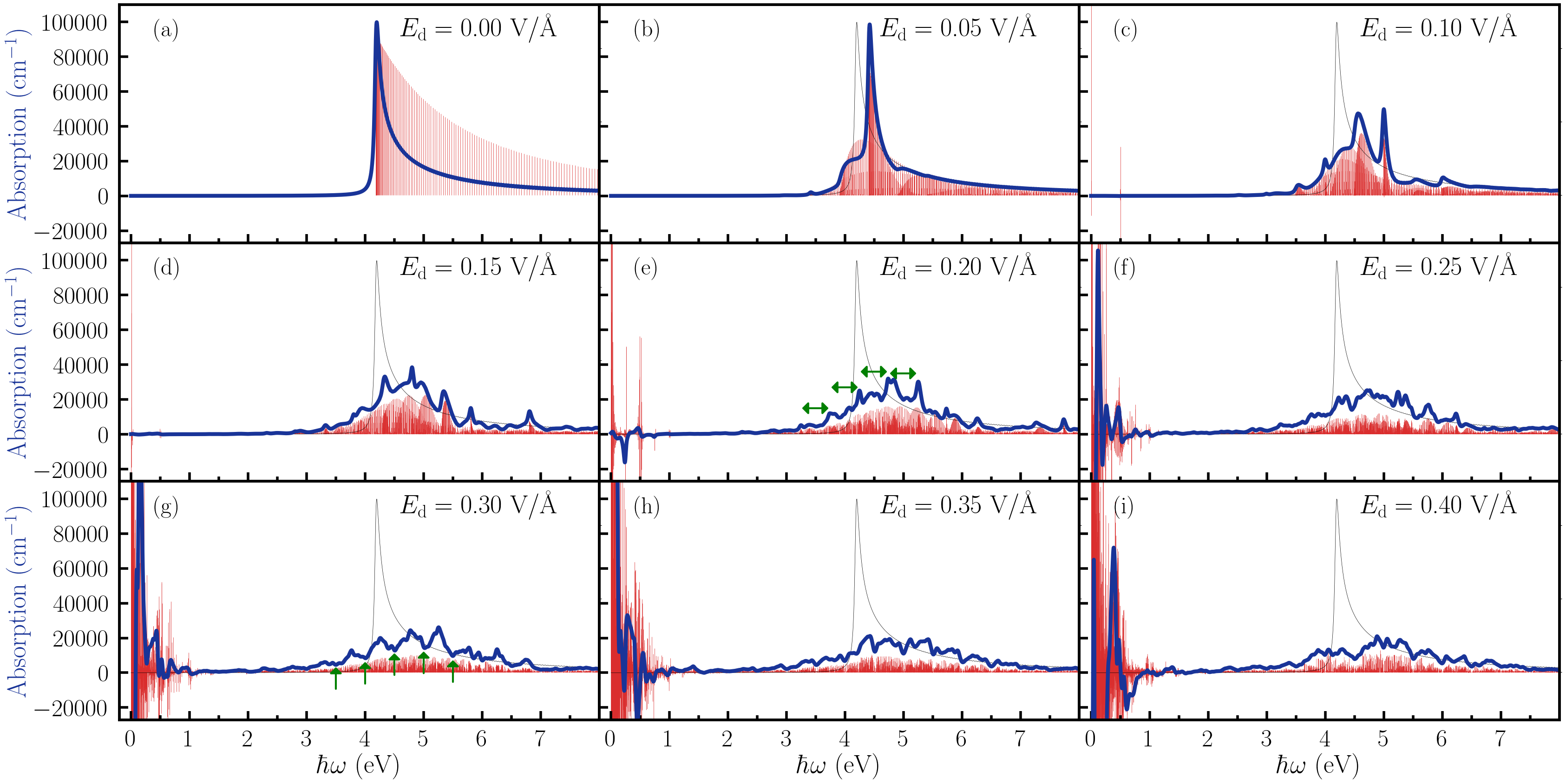}
\caption{Optical absorption spectrum of the laser-dressed cosine shaped lattice potential model as a function of the probe photon energy $\hbar\omega$. The panels (a)-(i) show how the spectra changes as the amplitude of the drive laser $E_{\textrm{d}}$ increases with drive photon energy $\hbar\Omega=0.5$ eV taken to be constant throughout. The red lines signals individual transitions. The blue lines represent the net absorption by broadening individual peaks with a Lorentzian function  with FWHM of $0.06$ eV. The gray line is the absorption spectrum for $E_{\textrm{d}}=0$ V/\r{A}.  In (e) green arrows each of width 0.5 eV are used to show the equidistant features in the laser-dressed spectra. Green arrows in (g) indicate the replicated dips occurring at integer multiples of 0.5 eV.}
\label{absmult}
\end{center}
\end{figure*}

Figure \ref{absmult} shows the laser-dressed optical absorption spectra of the model for different drive laser amplitude $E_{\textrm{d}} \in [0,0.4]$ V/\r{A}. The individual transitions are shown as red lines. These transition are broadened by a Lorentzian function with full width at half maximum (FWHM) of $0.06$ eV to yield the net absorption profile shown in blue.

The field-free absorption spectrum Fig. \ref{absmult}(a) has a sharp band edge at $4.2$ eV corresponding to transition at the direct band gap at  the $\Gamma$ point. As the amplitude of the drive electric field is increased  Fig. \ref{absmult}(b)-(i), several distinct changes in the net absorption spectrum emerge: (i) a blue shift of the band edge; (ii) a reduction of the intensity in  the main absorption features for $\hbar\omega \in [4.2,5] $ eV; (iii) creation of below band-gap absorption near the band edge (4.2 eV); (iv) The appearance of several sharp peaks in the absorption spectrum at $E_{\textrm{d}} \in [0.1,0.2] $ V/\r{A} that are separated by $\hbar\Omega$ [green arrows in Fig. \ref{absmult}(e)]; (v) Low-frequency ($\hbar\omega < 0.6$ eV) intense absorption and stimulated emissions for $E_{\textrm{d}} \in [0.25,0.4]$ V/\r{A}; and (vi) replicated dips in absorption spectrum [green arrows in Fig. \ref{absmult}(g)] for $E_{\textrm{d}} > 0.3 $ V/\r{A} at probe energy exactly equal to integer multiples of 0.5 eV. Below we discuss the origin of these changes. Overall, the one-dimensional solid, which in pristine form absorbs in the $\hbar\omega \in [4.2,8]$ eV range, after driving with non-resonant light becomes an absorber in $\hbar \omega \in [0,12]$ eV range. Thus, strong fields are seen to reversibly transform a semiconductor with a wide band gap into a broad band absorber!

\subsubsection{Blue shift of the band edge}\label{3a}

\begin{figure}[htbp]
\begin{center}
\includegraphics[width=0.5\textwidth]{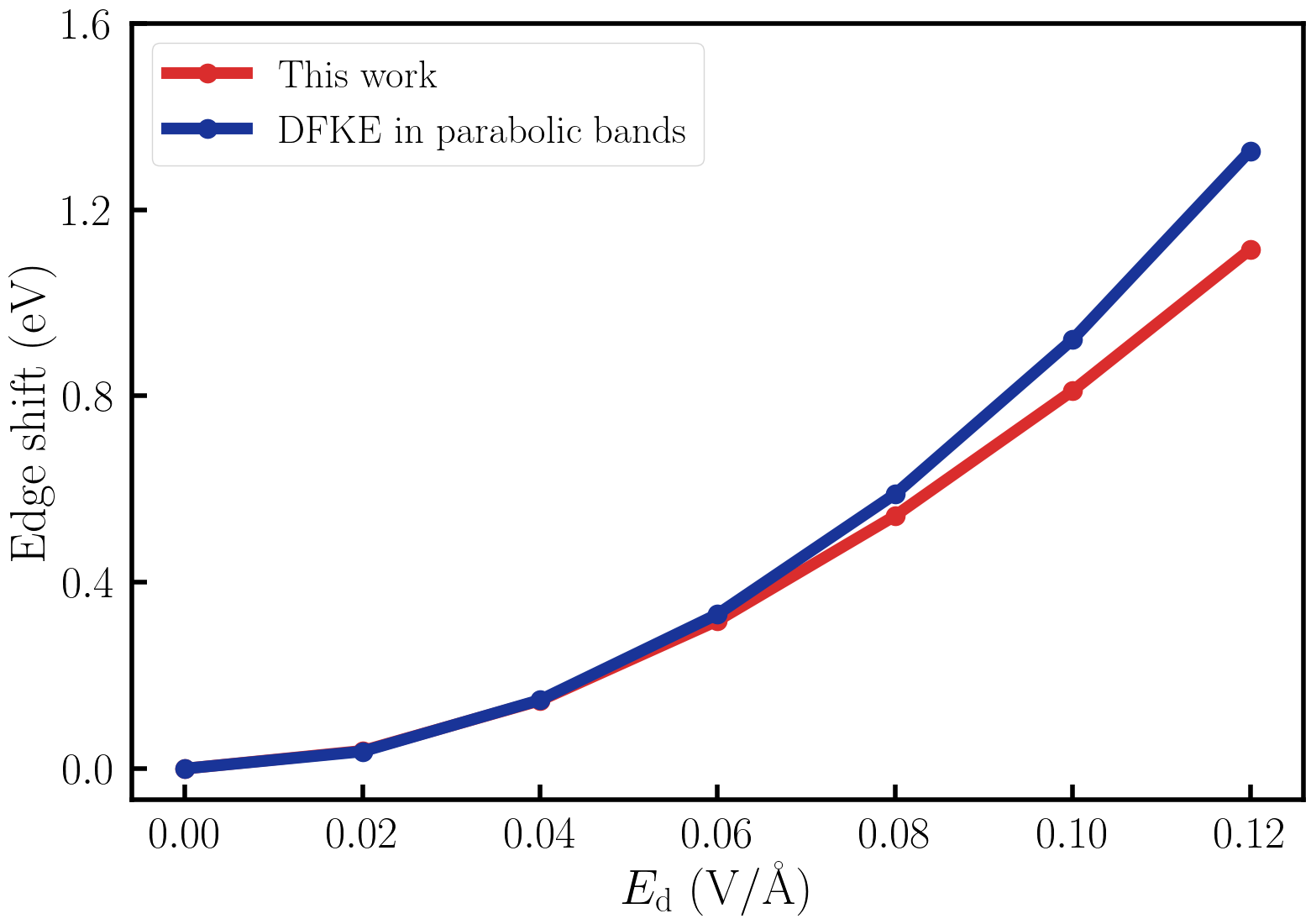}
\caption{Shift of the field-free band edge in the laser-dressed absorption spectrum as a function of the drive electric field amplitude. Red line: shift obtained from Eq. \eqref{final}. Blue line: shift predicted for a parabolic band model equal to the pondermotive energy \cite{Jauho1996,Johnsen1998}.}
\label{absshift}
\end{center}
\end{figure}

We first show that the theory quantitatively recovers the DFKE \cite{Jauho1996}. The effect is best known for a parabolic band model where the laser-driving  blue shifts the band edge by the pondermotive energy $U_{p}=\frac{e^2 E_{\textrm{d}}^2}{4m^{*}\Omega ^2}$ (where $m^{*}$ is the effective mass for the parabolic band model), creates below band-gap absorption and absorption sidebands. In our calculation, we define the band edge as the transition energy with maximum absorption strength occurring at the $\Gamma$ position. To compare this with DFKE predictions, we apply a two-band parabolic approximation to our model by calculating an effective mass as $\frac{1}{m^{*}}=\frac{1}{m_{\textrm{c}}}-\frac{1}{m_{\textrm{v}}}$, where $m_{\textrm{c,v}}=\left( \left. \frac{d^2 E_{\textrm{c,v}}(k)}{dk^2} \right\rvert_{k=0}\right) ^{-1}$ is the effective mass of conduction (c) and valence (v) band near the Fermi energy at the $\Gamma$ point and $E_{\textrm{c,v}}(k)$ the band energy dispersion \cite{haugbook}. We obtain an effective mass $m^{*}=0.082 m_{e}$ for the one dimension model here.

Figure \ref{absshift} shows the band edge shift from Eq. \eqref{final} (in red) and compares it with the DFKE prediction (in blue) for varying amplitudes of the drive electric field. Equation \eqref{final} recovers the DFKE results for $E_{\textrm{d}} \le 0.06$ V/\r{A}. The deviations for higher electric field amplitudes arise due to the nonparabolicity of the model and the presence of other bands, which are not included in the DFKE theory \cite{Jauho1996,Johnsen1998}. The observed band edge shift in Fig. \ref{absshift} can be understood in the  Floquet-picture through so-called Floquet-Bloch shifts \cite{Dimitrovski2017} of the bands at $\Gamma$ position, which are reminiscent to the repulsion of two-level systems under applied electric fields.

\subsubsection{Optical signatures of Floquet replicas} \label{3b}

\begin{figure*}[htbp]
\begin{center}
\includegraphics[width=\textwidth]{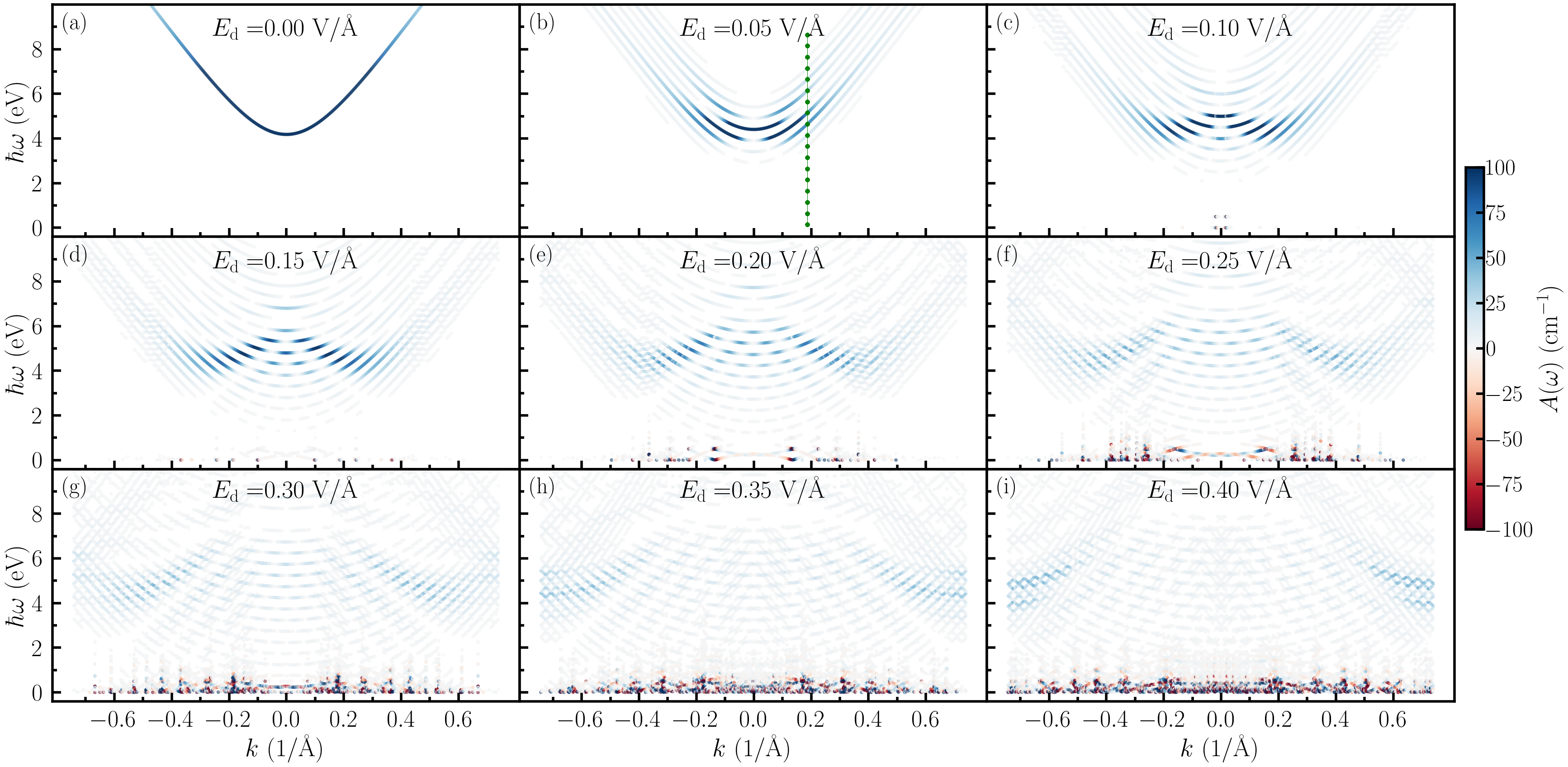}
\caption{Optical transitions responsible for the field dressed absorption spectra shown as function of crystal momentum $k$ on the $x$ axis and probe photon energy $\hbar\omega$ on the $y$ axis. The color blue in the heat map represents absorption and red stimulated emission. The vertical green line in (b) represents transitions happening at a fixed $k=0.187$ \r{A}$^{-1}$. The dots on the line indicates possible transitions occurring at equidistant energies separated by $0.5$ eV equal to the drive photon energy. Complicated structure arises at $\hbar \omega < 0.6$ eV for higher electric field amplitude [(f)-(i)] where a plethora of intense absorption and stimulated emission emerge as seen in Fig. \ref{absmult}. }
\label{absksp}
\end{center}
\end{figure*}

Figure \ref{absmult} show that the laser-dressing leads to the emergence of below band-gap absorption features, and characteristic peaks that are replicated at integer multiples of the drive photon energy [see green arrows in Fig. \ref{absmult}(e)]. These replicated absorption sidebands had been predicted in finite nanostructure \cite{Gu2018}. Strikingly, similar features are clearly visible here for the laser-dressed solid even in the presence of spectral congestion of the pristine absorption spectrum.

To understand the origin of these novel effects, consider Fig. \ref{absksp} where the contributions by the different $k$ points to the overall spectra are resolved. The heat maps signal the strength of absorption (in blue) and stimulated emission (in red). The range of the absorption coefficient $\in [-100,100]$ cm$^{-1}$ is chosen to enhance the visibility of the transitions in Fig. \ref{absksp}. The field-free  spectrum [Fig. \ref{absksp}(a)] shows the equilibrium absorption occurring throughout the Brillouin zone. It has a U-shaped structure as a result of the increase in gap between the highest valence and lowest conduction band when moving towards the edges of the Brillouin zone. Other possible transitions are of higher energy than the range of $\hbar\omega$ chosen in the figure. As the electric field amplitude increases, e.g., $E_{\textrm{d}}=0.05$ V/\r{A}, we observe the emergence of replicas of the field-free U-shaped structure both above and below in probe photon energy. The one that are below, contribute to the below band gap absorption. We draw a vertical green line in Fig. \ref{absksp}(b) at $k=0.187$ \r{A}$^{-1}$ with dots equally spaced by exactly the drive photon energy $\hbar\Omega=0.5$ eV. As seen, the dots lie exactly on the transitions at the same $k$ point. This indicates that the optical transitions are separated by integer multiples of drive photon energy. This is due to transitions happening among the same pair of Floquet-Bloch modes but across different FBZ. We can extend a similar analysis to all the $k$ points in the Brillouin zone, which gives rise to the replicated U-shaped structures in the figure. In fact, as the drive electric field amplitude is increased, we observe more and more of such replicated transitions giving rise to multiple copies of the U-shaped structure, which indicates that more Floquet-Bloch modes from different FBZ are participating in the net absorption spectrum leading to the replicated features seen in Fig. \ref{absmult}(e). The figure also shows that simple analyses based on the density of states in the Floquet-driven system \cite{Jauho1996, Johnsen1998}  are not enough to capture the nonequilibrium absorption spectra. Variations in the population factor and MME with $E_{\textrm{d}}$ and $k$ in the Brillouin zone, and  possible selection rules at play~\cite{Engelhardt2021}, lead to important additional structure as reflected by the interference like patterns in Fig. \ref{absksp}.

\subsubsection{Intense low-frequency transitions}\label{3c}

We now show that the hybridization of Floquet-Bloch modes leads to the opening of previously forbidden low frequency transitions with strong optical absorption features, and to dips in the absorption spectrum at integer multiples of the drive photon energy. For this discussion , it is useful to assess the interplay of the population factors and the Fourier components of the MME among the Floquet-Bloch modes leading to a net absorption signal. Using Eq. \eqref{final} we define the net absorption intensity for a transition occurring between  $\alpha$ and   $\beta$ Floquet-Bloch modes with $n$ FBZ separation at $\hbar\omega=E_{\alpha k}-E_{\beta k}+n\hbar\Omega$ as 
\begin{equation}
\label{tranint}
\Pi_{\alpha\beta k}^{(n)}=|\mathcal{P}_{\alpha\beta k}^{(n)}|^2(\Lambda_{\alpha\beta k}-\Lambda_{\beta\alpha k}) .
\end{equation}
The first term in Eq. \eqref{tranint} represents absorption while the second term represents stimulated emission of a photon. 

\begin{figure}[htbp]
\begin{center}
\includegraphics[width=0.5\textwidth]{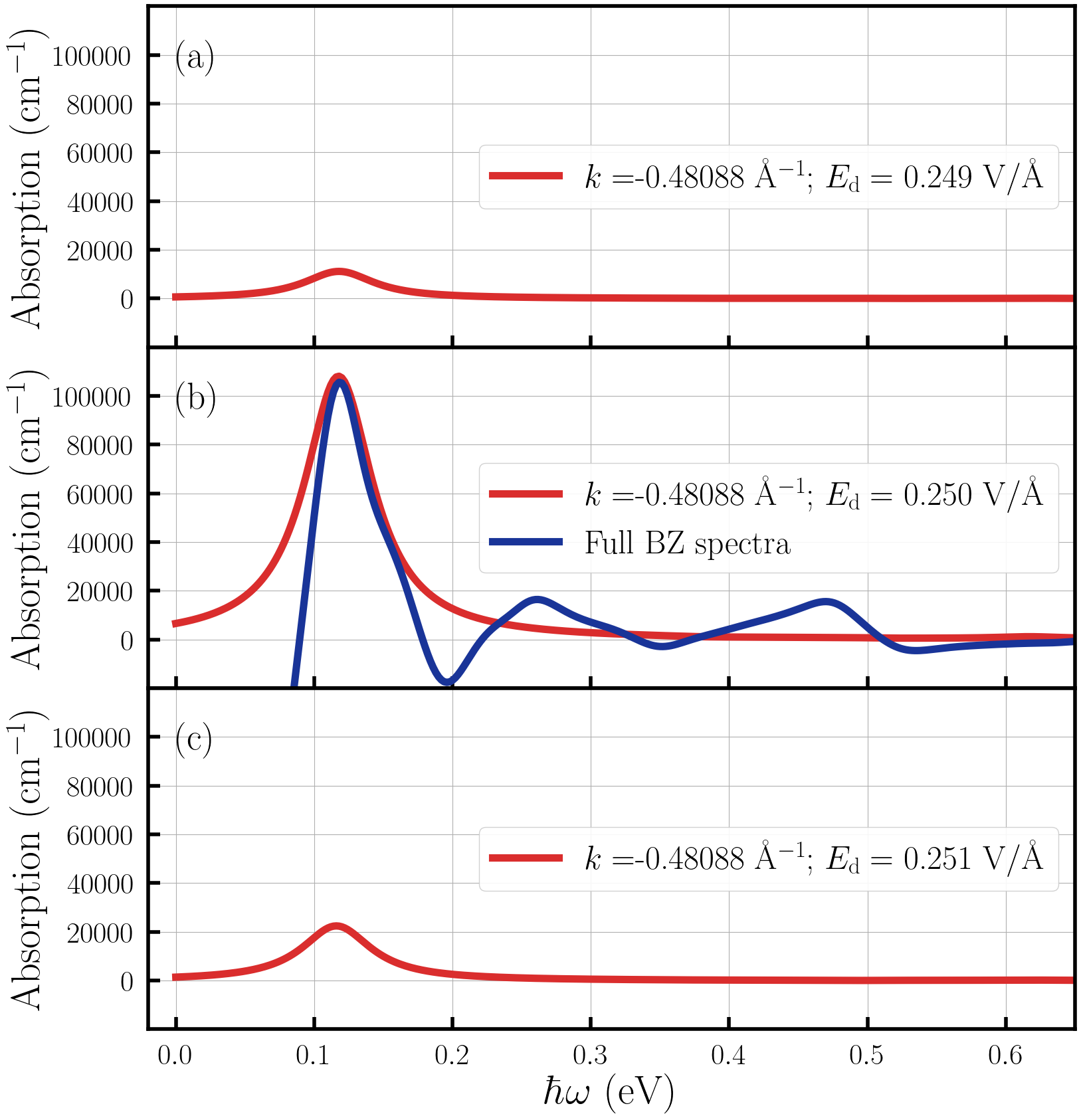}
    \caption{Net absorption spectrum for transitions occurring at $k=-0.48088$ \r{A}$^{-1}$ showing the emergence of intense low-frequency absorption feature due to hybridization of the Floquet-Bloch modes. The field-dressed absorption spectrum is plotted for drive electric field amplitude (a) 0.249 V/\r{A}; (b) 0.250 V/\r{A}; and (c) 0.251 V/\r{A}. Additionally in (b) we compare with the net absorption spectrum (blue line).
    }
\label{abshybdabs}
\end{center}
\end{figure}

\begin{figure}[htbp]
\begin{center}
\includegraphics[width=0.5\textwidth]{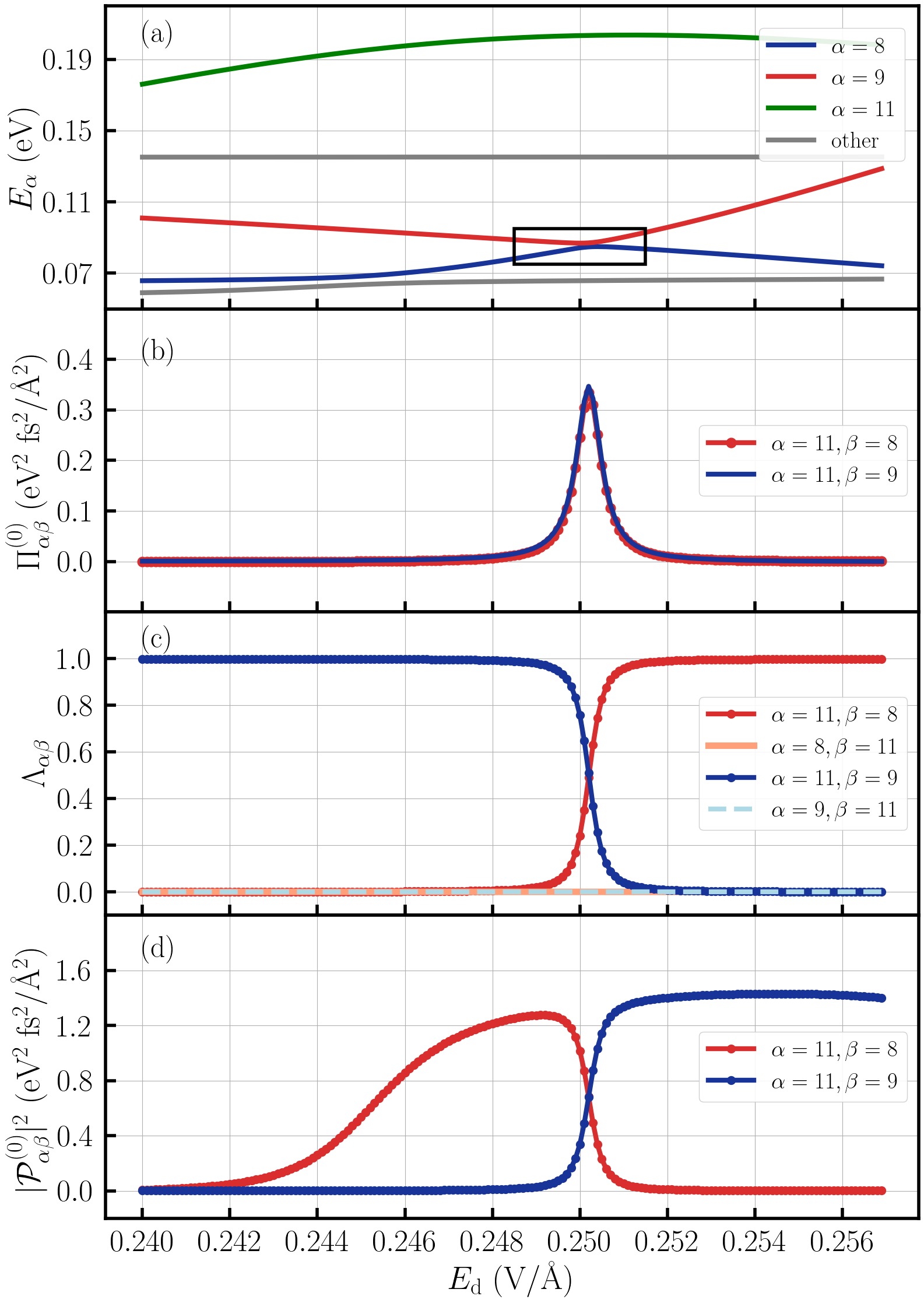}
    \caption{Hybridization of the Floquet-Bloch mode 8 and 9 at $k=-0.48088$ \r{A}$^{-1}$ leading to intense optical transition to Floquet-Bloch mode 11. (a) Quasienergies in the fist FBZ; (b) Absorption intensity $\Pi_{\alpha\beta k}^{(n)}$ [Eq. \eqref{tranint}]; (c) Population factor; and (c) Intra-FBZ MME for the participating Floquet Bloch modes as a function of $E_{\textrm{d}}$. The intense absorption feature arising due to this hybridization at $k=-0.48088$ \r{A}$^{-1}$ can be seen in the spectra [Fig. \ref{abshybdabs}(b) (red line)]  at $\hbar \omega=0.119$ eV for $E_{\textrm{d}}=0.25$ V/\r{A}.
    }
\label{abshybd}
\end{center}
\end{figure}

Figures \ref{absmult}(e)-(i) show intense absorption and stimulated emissions features in the field-dressed absorption spectra in the low-frequency region $(\hbar\omega < 0.6$ eV). These features are more intense than other features present at higher frequencies. We investigate the origin of these low-frequency transitions by isolating one particular feature and track how it changes upon varying the drive field amplitude in Fig. \ref{abshybdabs}. We plot the net absorption spectrum at $k=-0.48088$ \r{A}$^{-1}$  in Fig. \ref{abshybdabs} (a) for $E_{\textrm{d}}=0.249$ V/\r{A}; (b) for $E_{\textrm{d}}=0.25$ V/\r{A}; and (c) for $E_{\textrm{d}}=0.251$ V/\r{A} (red lines). We also plot the net absorption spectrum from the full Brillouin zone from Eq. \eqref{final} for $E_{\textrm{d}}=0.25$ V/\r{A} in Fig. \ref{abshybdabs}(b) (blue line). Both the spectra in Fig. \ref{abshybd}(b) essentially coincide suggesting that the intense absorption feature at $\hbar\omega=0.12$ eV arises due to transitions at $k=-0.48088$ \r{A}$^{-1}$. 

The Floquet-Bloch modes are  denoted by labels 1 through 11 in ascending order with the  quasienergies. The intense absorption feature seen at $k=-0.48088$ \r{A}$^{-1}$ is the result of  a transition  from Floquet-Bloch mode 8 to 11  with $\hbar\omega=0.119$ eV and 9 to 11 with $\hbar\omega=0.116$ eV in the same FBZ. As shown the magnitude of the  absorption feature diminishes rapidly as the drive field amplitude is changed from $E_{\textrm{d}}=0.25$ V/\r{A}.

In Fig. \ref{abshybd} we investigate the factors contributing to the low-frequency transition at $E_{\textrm{d}}=0.25$ V/\r{A} and its strong dependence on $E_{\textrm{d}}$. Figure \ref{abshybd}(a) shows the quasienergies of the modes 8, 9 and 11 in the first FBZ, (b) shows the changes in absorption intensity $\Pi_{11,8 k}^{0}$ and $\Pi_{11,9k}^{0}$, (c) the population factors; and (d) the intra-FBZ MME between the participating modes as a function of $E_{\textrm{d}}$. As seen, the quasienergies for mode 8 and 9 (black box)  form an avoided crossing at around $E_{\textrm{d}}=0.25$ V/\r{A}. 

Figures \ref{abshybd}(b)-(d) show sudden changes that coincide with the emergence of the intense absorption feature. Away from the hybridization zone, there is not net intra-FBZ optical absorption between mode 8 and 11 or 9 and 11 ($\Pi_{11,8k}^{0}$=$\Pi_{11,9k}^{0}$=0). This is because for the states that have non-zero intra-FBZ MME (see e.g., $|\mathcal{P}_{11,8k}^{(0)}|^2$ for $E_{\textrm{d}}$  before the avoided crossing), do not have favorable population factors ($\Lambda_{11,8k}$=$\Lambda_{8,11k}$=0). Alternatively, if they have favorable population factors (such as $\Lambda_{11,8k}$ after the avoided crossing), they have zero intra-FBZ MME (see 
 $|\mathcal{P}_{11,8k}^{(0)}|^2$). Only for $E_{\textrm{d}}$ around the avoided crossing, the population factor and intra-FBZ MME changes in such a way that it opens a strong $\Pi_{11,8k}^{0}$ and $\Pi_{11,9k}^{0}$ [Fig. \ref{abshybd}(b)] and an intense absorption feature seen. That is, the hybridization of Floquet-Bloch modes leads to the opening of previously forbidden transitions with strong optical absorption features. This hybridization between Floquet-Bloch modes can open channels of either absorption or stimulated emission leading to a plethora of intense low-frequency features in the absorption spectrum [Fig. \ref{absmult}(e)-(i)]. These transitions can be to other Floquet-Bloch modes  in the FBZ as shown in Fig. \ref{abshybd} or between the two modes involved in the hybridization.

\subsubsection{Replicated dips in the absorption spectra}

\begin{figure}[htbp]
\begin{center}
\includegraphics[width=0.5\textwidth]{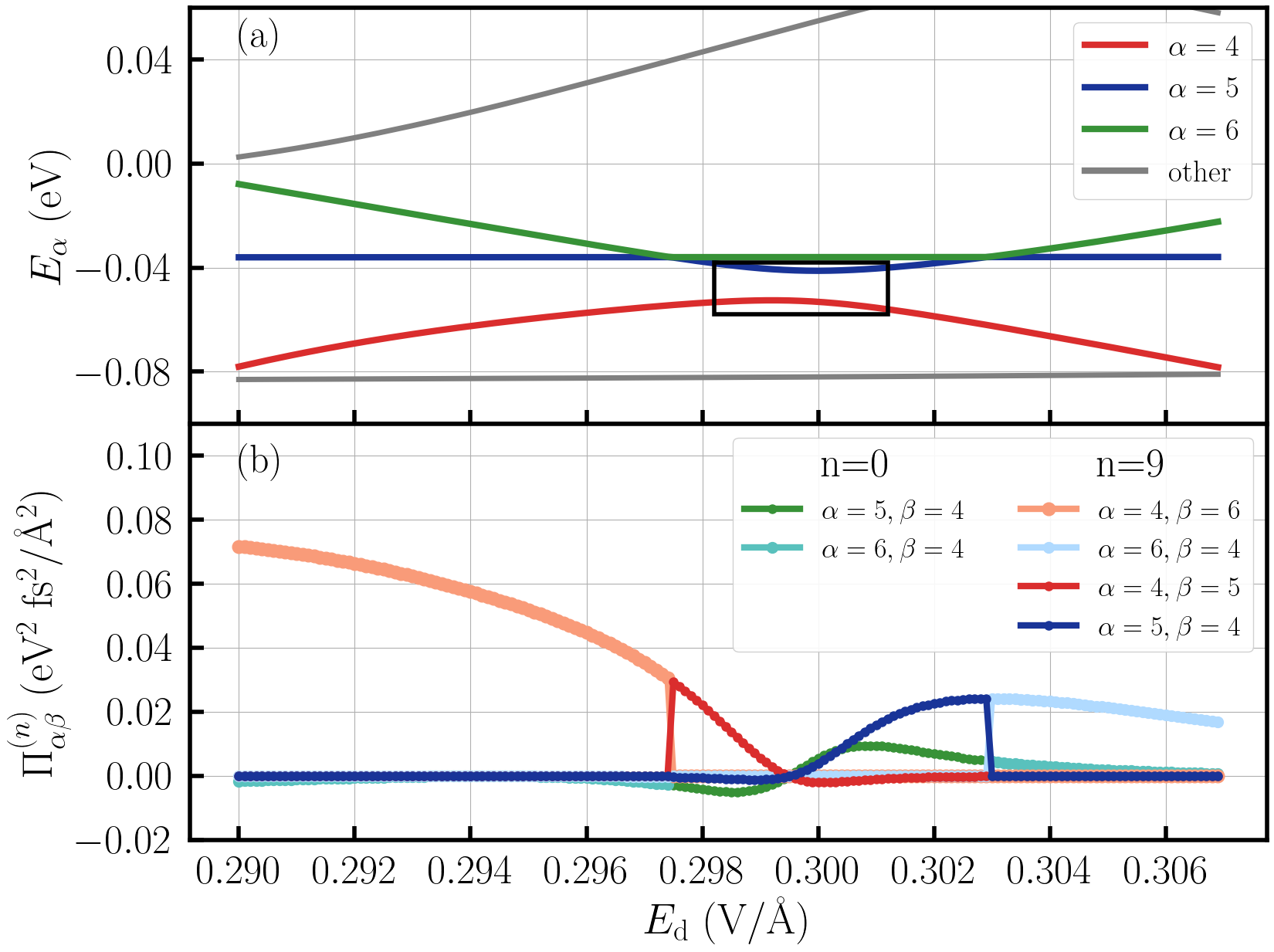}
\caption{Hybridization of Floquet-Bloch modes leading to suppression of net optical absorption. (a) Quasienergies of Floquet-Bloch modes 4,5 and 6 at $k=-0.43635$ \r{A}$^{-1}$ as a function of $E_{\textrm{d}}$ showing the formation of avoided crossing due to hybridization of mode 4 and 5 (black box). (b) Absorption intensity among modes 4 ,5 and 6 for transition with $n=0$ FBZ separation and $n=9$ FBZ separation as a function of drive field amplitude. Note the suppression of the absorption intensity of both 0 and 9 FBZ separation among mode 4 and 5 around the avoided crossing.}
\label{abshybdgaps}
\end{center}
\end{figure}

In addition to the replicated peaks, the absorption spectra in Fig. \ref{absmult}(g)-(i) also features replicated dips [green arrows in Fig. \ref{absmult}(g)] that are present at probe frequency equal to integer multiples of $\hbar\Omega=0.5$ eV. The dips are similar to observed gap openings in tr-ARPES spectrum \cite{Wang2013} or dips in optical conductivity of graphene \cite{Zhou2011} and are known to arise due to creation of gaps in the quasienergies from the hybridization of the Floquet states. 

We address the formation of replicated dips using the example of a transition occurring around one such dip observed in the spectrum at 4.5 eV for $E_{\textrm{d}}=0.3$ V/\r{A} in Fig. \ref{absmult}(g). This optical feature involves a transition between Floquet-Bloch mode 4 and 5 at $k=-0.43635$ \r{A}$^{-1}$ with a 9 FBZ separation. We plot the quasienergy of modes 4, 5 and 6 in Fig. \ref{abshybdgaps}(a) as a function of drive field amplitude. As seen, modes 4 and 5 form an avoided crossing (black box) around $E_{\textrm{d}}=0.3$ V/\r{A}. Since transitions among these two modes can happen across $n$ FBZs, the avoided crossings effectively creates a dip in the spectrum when  $\hbar\omega \approx n\hbar\Omega$. In Fig. \ref{abshybdgaps}(b) we plot the net absorption intensity as a function of drive field amplitude for transitions that can happen within the FBZ ($n$=0) and transition with 9 FBZ separation ($n$=9) among the modes 4 and 5. Figure \ref{abshybdgaps}(b) shows the transitions $4\rightarrow 5$ and $5\rightarrow 4$ give rise to features around 0.0 eV and 4.5 eV. Around the avoided crossing, the hybridization of modes 4 and 5 lead to changes in the $n=0$ transition intensity. It lead to  stimulated emission just before the crossing and absorption just after the crossing creating low-frequency transitions in the absorption spectrum.  However, $\Pi_{5,4k}^{(9)}$ and $\Pi_{4,5k}^{(9)}$ vanish at the avoided crossing $E_{\textrm{d}}=0.2993$ V/\r{A} around  $E_{\textrm{d}}=0.3$ V/\r{A} leading to no net transition at probe energy $9\hbar\Omega=4.5$ eV. Similarly the transition intensity $\Pi_{5,4k}^{(0)}$ vanish around the avoided crossings leading to dip around 0.0 eV. Hence, at $k=-0.43635$ \r{A}$^{-1}$ the hybridization of mode 4 and 5 leads to dips in absorption spectrum at integer multiples $n\hbar\Omega$. 

The dips in the absorption spectrum at $\hbar\omega = n\hbar\Omega$ occur whenever any two  modes for which transitions are allowed hybridize. For this reason, the effect survives the congested transitions throughout the Brillouin zone and is clearly visible in the absorption spectrum in Figs. \ref{absmult}(g)-(i).

\section{Conclusions}\label{sec5}

\begin{figure*}[htbp]
\begin{center}
\includegraphics[width=\textwidth]{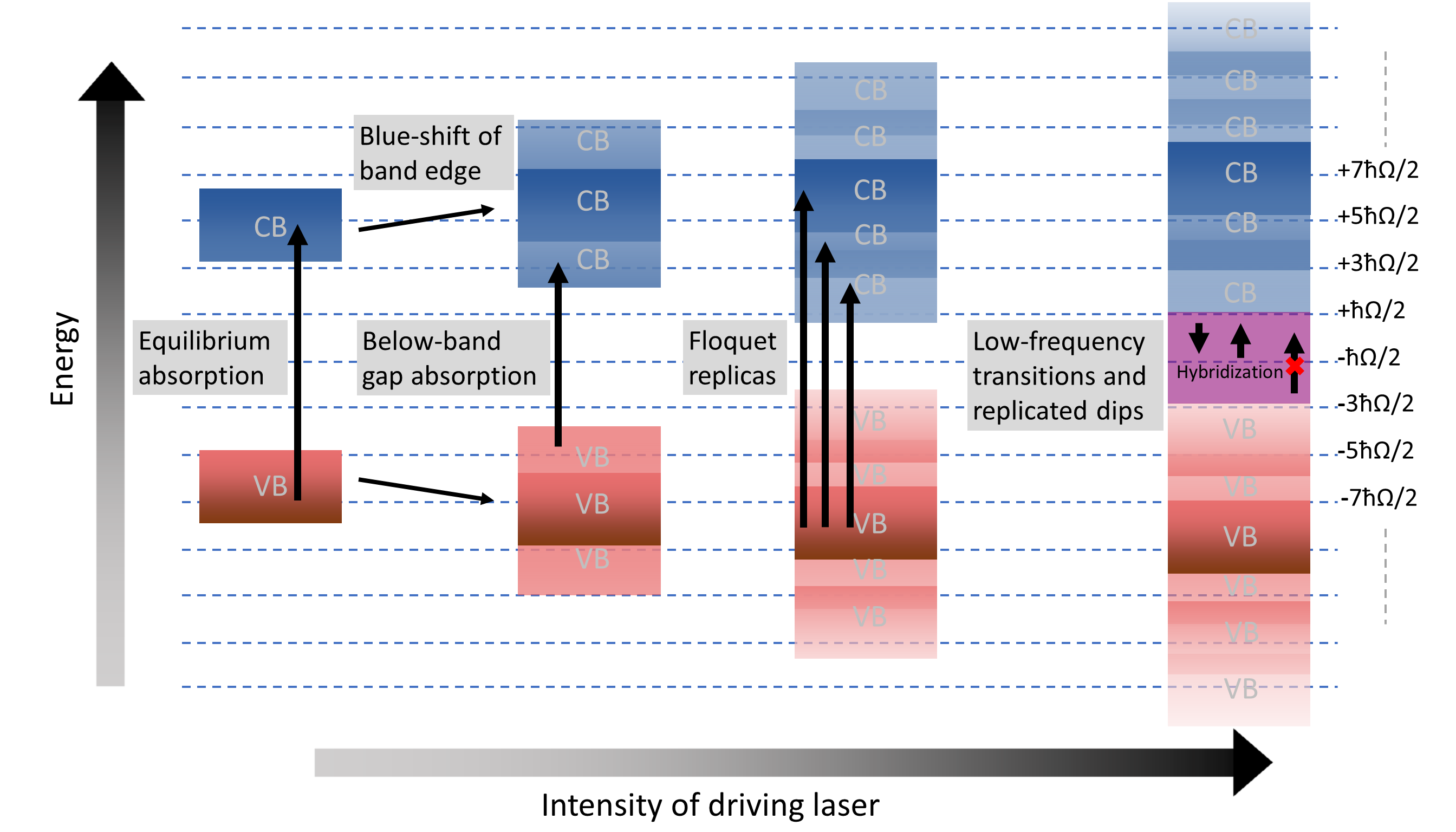}
    \caption{Schematic representation of the various phenomena that occur in the optical absorption of laser-dressed solids. Our theory shows that the Floquet-Bloch modes are the natural states to interpret laser-dressing of solids as the net optical absorption can be understood as transitions between these modes across different number of Floquet-Brillouin zones. The below band-gap absorption transitions are the lower energy replicas of the transitions below the band edge. Replicated features seen in the absorption spectrum are a result of transitions occurring among the modes separated by successive FBZ. Hybridization of Floquet-Bloch modes within the FBZ lead to intense low-frequency transitions, and dips in the spectrum at integer multiples of drive photon energy. }
\label{abssummary}
\end{center}
\end{figure*}

In conclusion, we have developed a general theory and computational strategy that now enables modeling and interpreting  the linear optical absorption of solids that are dressed by light of arbitrary strength and photon energy. The theory applies to crystalline solids of any band structure and dimensionality. In the theory, the effect of the driving laser is treated exactly using Floquet theory while the influence of the probing light is captured to first order in perturbation theory. 
In this context we isolated an expression [Eq. \eqref{final}] for the absorption coefficient of the laser-dressed solid that can be calculated via diagonalization techniques in Sambe space, thus avoiding explicitly propagating in time the dynamics of the material in the presence of the driving and probing laser. 

Remarkably, the resulting formula for net absorption is akin to the equilibrium theory of optical absorption but with the Floquet-Bloch modes playing the role of pristine eigenstates. That is, the non-equilibrium absorption properties in the laser-dressed solids arise due to transitions among Floquet-Bloch modes across several  Floquet-Brillouin zones. The Bohr-transition energies are determined by the difference between the quasienergies of the involved Floquet-Bloch modes. The transition strength is determined by a population factor which guarantees that the initial state is occupied and the final state is empty, and the Fourier components of the momentum matrix elements among the participating Floquet-Bloch modes. Both these quantities are dependent on the the driving laser. The theory was  implemented in a general purpose code that is available through \texttt{GitHub} \cite{code} and that can be interfaced with standard first-principle based electronic structure codes.

To isolate the emergent phenomenology due to strong laser-dressing, we computed the nonequilibrium absorption spectrum of a one-dimensional solid with cosine-shaped lattice potential. Overall, the laser-dressing transiently transforms a wide band-gap semiconductor into a broadband absorber. We demonstrated that the theory naturally recovers the dynamical Franz-Keldysh effect (DFKE) and showed that the pondermotive shift predicted by DFKE theory is only valid for relatively weak laser driving amplitudes (of $E_0 \leq 0.06$ V/\AA~ in our problem). In addition, the theory predicts additional important changes  in the absorption spectrum upon dressing with light that are beyond the scope of DFKE. 

We summarize our findings of the various phenomena observed in laser-dressed solids in Fig. \ref{abssummary}. For pristine matter, the probe laser induces vertical transitions between the valence (shown in red) and conduction (shown in blue) bands, leading to the equilibrium absorption spectrum. When the system is driven by a laser it leads to the formation of Floquet-Bloch modes \cite{Wang2013,Park2022,Ito2023}. This also results in the shifting of the pristine valence and conduction bands in such a way that it leads to a net blue-shift of band edge, akin to the Stark and Bloch-Seigert shift \cite{Sie2017, Dimitrovski2017}. Transitions induced among the newly formed Floquet-Bloch modes are observed as below band gap features in the absorption spectrum. Increasing the drive laser field intensity creates access to additional Floquet-Bloch replicas. Transitions that happen among these replicas of the pristine bands lead to sidebands in the absorption spectrum. At even higher laser intensities, the Floquet-Bloch modes hybridize (shown in purple) which leads to intense low-frequency transitions and dips in the absorption spectrum. All features in the absorption spectrum can be explained through transitions induced by the probe laser among the Floquet-Bloch modes.

From this analysis we can identify three purely optical signatures of the Floquet-Bloch states. First, the transitions among the Floquet replicas lead to absorption sidebands. We find that these features can be evident even in the congested electronic structure characteristic of solids. In addition, the hybridization of Floquet-Bloch modes leads to the emergence of low-frequency features and also dips in the absorption spectra [\emph{cf.} Figs. \ref{absmult}(e-i)]. Absorption spectroscopy experiments should be able to observe  these signature features of the formation of Floquet-Bloch states.

This paper paves the way toward establishing the response properties of laser-dressed matter as needed to develop materials with on-demand properties. Future prospects include developing a theory for the nonlinear optical response, characterizing the influence of electronic correlations and electron-phonon interactions, capturing effects due to spatial variations of the laser field that go beyond the dipole approximation, introducing open boundary conditions as needed to develop a theory of lineshapes and electron transport in laser-driven matter \cite{Cabra2020}, and investigating the differences in the light-dressing that can be achieved with quantum and classical light.

\begin{acknowledgments}
This material is based upon work supported by the National Science Foundation under Grant No. CHE-2102386.
\end{acknowledgments}



\bibliography{floqtics_refs}


\end{document}